\documentclass[12pt]{article}
\pdfoutput=1
\usepackage{geometry}
\usepackage{xcolor}
\usepackage{a4}
\usepackage{alltt}
\usepackage{graphicx}
\usepackage{epsf}
\usepackage{amsmath}
\usepackage{amssymb}
\usepackage{cite}
\usepackage{hyperref}
\usepackage{blindtext}
\usepackage[toc,page]{appendix}

\newcommand{\be}{\begin{equation}}
\newcommand{\ee}{\end{equation}}

\newcommand{\Rmnum}[1]{\expandafter\@slowromancap\romannumeral #1@}
\newcommand{\bea}{\begin{eqnarray}}
\newcommand{\eea}{\end{eqnarray}}
\begin{document}
%%%%%%%%%%%%%%%%%%%%%%%%%%%%%%%%%%%%%%%%%%%%%%%%%%%%%%%%%%%%
\title{\bf Cosmological bounces in spatially flat FRW spacetimes
  in metric $f(R)$ gravity} 
\author{Niladri Paul$^\ddagger$\footnote{Niladri Paul, did most of the work 
presented in this article when he was in the Physics Department of Indian 
Institute of Technology, Kanpur 208016, India.}, Saikat Nil 
Chakrabarty$^{**}$, Kaushik Bhattacharya$^{**}$ 
\thanks{ E-mail:~ npaul@iucaa.ernet.in, snilch@iitk.ac.in, 
kaushikb@iitk.ac.in} \\
\normalsize $^\ddagger$ IUCAA, Post Bag 4, Pune University Campus, Ganeshkhind,\\
\normalsize Pune 411 007, India\\
\normalsize $^{**}$ Department of Physics, Indian Institute of Technology,\\ 
\normalsize Kanpur 208016, India}
\maketitle
%%%%%%%%%%%%%% 
\begin{abstract}
The present work analyzes the various conditions in which there can be
a bouncing universe solution in $f(R)$ gravity. In the article an
interesting method, to analyze the bouncing FRW solutions in a
spatially flat universe using $f(R)$ gravity models using an effective
Einstein frame description of the process, is presented. The analysis
shows that a cosmological bounce in the $f(R)$ theory need not be
described by an equivalent bounce in the Einstein frame description of
the process where actually there may be no bounce at all. Nevertheless
the Einstein frame description of the bouncing phenomena turns out to
be immensely important as the dynamics of the bounce becomes amenable
to logic based on general relativistic intuition. The theory of scalar
cosmological perturbations in the bouncing universe models in $f(R)$
theories has also been worked out in the Einstein frame.
\end{abstract}
%%%%%%%%%%%%%%%%%%%%%%%%%%%%%%%%%%%%%%%%%%%%%%%
\section{Introduction}

If the BICEP2 \cite{Ade:2014xna} results are confirmed by future
observational projects, as the Planck collaboration, then the
scientific community can be fairly certain that some form of
inflationary dynamics must be responsible for the growth of the very
early universe. Inflation as a theory \cite{Riotto:2002yw,
  Linde:1990ta} solves various problems in cosmology, as the horizon
problem, the entropy problem, the flatness problem and others. But
standard inflationary theories have some inherent problems in them as
the trans-Planckian problem \cite{Brandenberger:2000wr, Martin:2000xs,
  Brandenberger:2002ty, Martin:2002kt, Brandenberger:2012aj}. If
inflation indeed happened in the very early phase of the universe then
the earliest modes (which are entering our horizon now) which became
superhorizon had wavelengths which are smaller than the Planck length
at the beginning of inflation. If this happens it produces severe
problem in the inflationary paradigm because then the calculations
concerning the cosmological perturbations due to these modes, based on
known laws of physics, breaks down.  There are various propositions
which tries to address these issues.  Some authors have tried to
modify the dispersion relation of the earliest modes
\cite{Martin:2002kt} while others have proposed cosmological bounces
in the general relativistic framework
\cite{Liu:2013kea,Piao:2003zm} and some variations of it
\cite{Cai:2008qw,Cai:2009rd,Cai:2012va,Cai:2013kja,Bhattacharya:2013ut,
Cai:2011tc, Cai:2009in, Qiu:2013eoa}.
Except these ideas there are proposals to bypass Einstein gravity and
work with $f(R)$ gravity induced bouncing universe scenarios where the
initial singularity in the big-bang model never arises
\cite{Bamba:2013fha, Novello:2008ra, Carloni:2005ii}. Some interesting
work regarding the stability of $f(R)$ theories was done in
Ref.~\cite{Barrow:1983rx}. The $f(R)$ theory paradigm has also been
used to provide a unified platform for inflation and late time
acceleration of the universe \cite{Nojiri:2003ft}.  Except the physics
of the early universe $f(R)$ theories of gravity have been used in
various other cosmological sectors as well, as in the problem of late
time cosmic acceleration \cite{Carroll:2003wy, Carroll:2004de,
  Amendola:2006we, Nojiri:2006ri}, the dark matter problem
\cite{Cembranos:2008gj} and many others \footnote{Most of the
  applications of $f(R)$ theories are appropriately summed up in the
  review articles in, Ref.~\cite{Nojiri:2010wj},
  Ref.~\cite{Sotiriou:2008rp} and Ref.~\cite{DeFelice:2010aj}. The
  mathematically inclined reader may also like
  Ref.~\cite{Clifton:2007ih, Wands:1993uu} where some general features of $f(R)$
  theory are nicely presented.}. In the present article we will be
dealing with higher derivative $f(R)$ theories of gravity and try to
analyze the dynamics of the bouncing cosmologies for spatially flat
FRW spacetimes.

We assume that in the very early phase of the universe, when the Ricci
scalar may have attained a very high value (compared to some relevant
preassigned scale), the dynamics of the universe was guided by an
effective $f(R)$ theory of gravity which assisted a cosmological
bounce. This naturally leads to a form of second order $f(R)$ which
was used by Starobinsky \cite{Starobinsky:1980te} to address the issue
of inflation. In essence $f(R)$ theory enters our cosmological model
in the same way as in Starobinsky's model of inflation but unlike
Starobinsky's model where the modified gravity theory produces
inflation in a universe devoid of matter, in our case the modified
theory of gravity will ensure a non singular cosmological bounce in
presence or absence of hydrodynamic matter. As in $f(R)$ theories of
gravity the equations of motion of the metric involves higher order
partial derivatives of the metric (compared to the second order
theories in GR) we will often call $f(R)$ theories of gravity as
higher derivative (HD) theories of gravity.  In these HD theories of
gravity the study of cosmological bounces become difficult because of
the HD terms of the metric and consequently most of the times an
analytic solution remains illusive. To tackle this problem we have
proposed an alternative way of looking at these bouncing solutions in
HD theories of gravity. Our method relies on the description of the
bouncing phenomenon in a conformally related Einstein frame. Although
the relations connecting the conformal frame, where an HD theory of
gravity lives, and that of a conformally related Einstein frame were
well studied for the case of Starobinsky inflation
\cite{Starobinsky:1980te}, the present analysis of the bouncing
phenomena in the two frames requires a complete new way of
interpreting the physics of cosmological evolution. This is because
unlike the case of Starobinsky inflation here the conformally related
frames will not portray the same bouncing phenomenon when the spatial
curvature of the FRW spacetime vanishes. In this particular case if
one assumes that there is a cosmological bounce in the HD theory then
in the Einstein frame description of the events there is no equivalent
bounce. As a consequence of this, one has to specify from the first
which is the physical conformal frame and which is an auxiliary
conformal frame used mainly to solve the problem.

The Einstein frame description of cosmological bounces, in presence of
matter, requires a scalar field whose potential may not be bounded
from below and a hydrodynamic fluid whose energy density and pressure
is modulated by the scalar field strength. But these facts do not
deter one from describing the bouncing phenomenon in the Einstein
frame as because in the time period where the HD theory is supposed to
be active the scalar field cannot roll down to its negative infinite
potential depth. In our analysis of the bouncing phenomena in HD
theories we will assume that the hydrodynamic matter (if present) will
satisfy the weak energy conditions (WEC) at all times \footnote{The
  validity of the WEC naturally implies the validity of the null
  energy condition.}. If WEC is violated then many of our predictions
require to be modified. But fortunately it can be shown that in the HD
theory of gravity one can have perfectly non-singular cosmological
bounces for the FRW solutions with flat spatial hypersurfaces where
the hydrodynamic matter always satisfies WEC. In the Einstein frame
the energy conditions become intricate as there are two components of
matter: one hydrodynamic and the other comprising of a scalar field
whose potential is unbounded from below.

It has been shown in the article that the analysis of the bouncing
phenomenon in the Einstein frame gives us many insights about the
behavior of the cosmological evolution in the physical frame. Whether
we will have a symmetrical bounce or an asymmetrical bounce in the HD
theory can be predicted from the values of the scalar field and its
first derivative with respect to the time variable in the Einstein
frame at the time of bounce. Ultimately one has to use numerical
methods to calculate the time evolution of the system but the
numerical codes can be better handled in the Einstein frame. The
article ends with an analysis of cosmological perturbations in the
bouncing universe. The calculations are done in the Einstein frame.
Once the evolution of the perturbations are known in the Einstein
frame one can map the solutions back to the conformal frame where the
HD theory lives.

The material in the article is presented in the following way. The
next section describes the very general framework of HD theories of
$f(R)$ gravity and the bouncing conditions for spatially flat FRW
spacetimes. In section \ref{cp} we present the topic of relativity of
conformally connected frames which deals with the description of the
cosmological bounce phenomenon as observed, in the conformal frame
where the HD theory lives and in the related Einstein frame. Section
\ref{numb} describes the bouncing phenomenon in the HD theory where we
present the numerical results of actual calculations regarding
cosmological bounce in FRW spacetimes with flat spatial sections in
presence of radiation.  The next section \ref{einf} gives a detailed
description of the methods employed to solve the dynamical equations
in the Einstein frame. Section \ref{perts} addresses the issue of
scalar cosmological perturbations in the bouncing universe models as
observed from the Einstein frame. The article ends with a discussion
and summary of the results presented in it. 
%%%%%%%%%%%%%%%%%%%%%%%%%%%%%%%%%%%%%%%%%%%%%%%%%%%%%%%%%%%%%
\section{The cosmological model}
\label{cmodel}

In this article we will mainly focus on bouncing cosmologies in the
spatially flat FRW universe.  In GR one cannot have a cosmological
bounce in the flat FRW universe when matter obeys standard energy
conditions \cite{Martin:2003sf}. On the other hand in higher
derivative gravity one can have bouncing cosmologies in flat FRW case,
as reported in Refs.~ \cite{Bamba:2013fha, Novello:2008ra}.  In this
article we will focus on the forms of $f(R)$ as given by
\begin{eqnarray}
f(R)=R+\alpha  R^{n}\,,
\label{ngrav}
\end{eqnarray}
where $\alpha$ is a real number and $n$ is an integer greater than
one.  In Starobinsky's model of inflation $n=2$ and $\alpha>0$. The
major portion of this article will be dealing with the case where
$n=2$ but unlike Starobinsky's model one requires $\alpha<0$ for a
cosmological bounce. This form of $f(R)$ will modify the early stages
of evolution of the universe and can only produce a cosmological
bounce in presence of matter. The correction term in $f(R)$ which is
$\alpha R^n$ may have originated from quantum corrections
\cite{Utiyama:1962sn} to the standard GR action, in the early
universe. Before going into the formalism of the HD theory of gravity
predicted by the quadratic form of $f(R)$ the reader must be reminded
that the quadratic theory of gravity with a negative $\alpha$ is by
itself not stable as one cannot ensure $f'(R)>0$ throughout the cosmic
evolution. This problem can be avoided in certain cases, where the
instability can be avoided by suitably choosing the parameter
$\alpha$, or by generalizing the form of $f(R)$ in such a way that the
instabilities are taken care of. As this topic requires a detailed
discussion we present it in a later subsection \ref{stab}, after presenting the basic
formalism of HD theories.

Following the notation of Ref.~\cite{Sotiriou:2008rp}, the fundamental
equation of $f(R)$ theory which corresponds to the 
Einstein's equation in GR, is
\begin{eqnarray}
G_{\mu\nu}\equiv R_{\mu\nu}-\frac{1}{2}g_{\mu\nu}R &=&\frac{\kappa T_{\mu\nu}}
{f^{\prime}(R)}+g_{\mu\nu}\frac{\left[f(R)-Rf^{\prime}(R)\right]}{2f^{\prime}(R)}
\nonumber\\
&+&\frac{\nabla_{\mu}\nabla_{\nu}f^{\prime}(R)-g_{\mu\nu}\square f^{\prime}(R)}
{f^{\prime}(R)}\,,
\label{ein1}
\end{eqnarray}
where the prime denotes a derivative with respect to the Ricci scalar
$R$, $T_{\mu\nu}$ is the energy momentum tensor of the fluid which
permeates the spacetime, $\nabla_\mu$ is the covariant derivative and
$\square \equiv \nabla^\mu \nabla_\mu$. Using the FRW line element
\begin{eqnarray}
ds^2 = -dt^2 + a^2(t)\left[\frac{dr^2}{1-kr^2} + r^2(d\theta^2 +
  \sin^2 \theta\,\,d\phi^2)\right]\,,
\label{frw}
\end{eqnarray}
and the energy momentum tensor of a perfect fluid
\begin{eqnarray}
T_{\mu \nu} = (\rho + P)u_\mu u_\nu + P g_{\mu\nu}\,,
\label{tmunu}
\end{eqnarray}
one can write the various components of Eq.~(\ref{ein1}) in the form
\begin{eqnarray}
3\left(\frac{k}{a^2}+H^2\right) &=&\frac{\kappa}{f^{\prime}(R)} (\rho
  + \rho_{\rm eff})\,,
\label{fried}\\
\frac{k}{a^{2}}+3H^{2}+2\dot{H} &=&\frac{-\kappa}{f^{\prime}(R)}
(P+ P_{\rm eff})\,,
\label{2ndeqn}
\end{eqnarray}
where $H$ is the conventional Hubble parameter defined as $H\equiv
\dot{a}/a$ and $k$ stands for the constant specifying the curvature of
the 3-dimensional spatial hypersurface. The dot specifies a derivative
with respect to cosmological time $t$ and $\rho_{\rm eff}$ and $P_{\rm
  eff}$ are defined as:
\begin{eqnarray}
\rho _{\rm eff} &\equiv& \frac{Rf^{\prime}-f}{2\kappa}-\frac{3H\dot{R}
f^{\prime \prime}(R)}{\kappa}\,,
\label{reff}\\
P_{eff} &\equiv& \frac{\dot{R}^{2}f^{\prime \prime \prime} + 2H\dot{R}f^{\prime
    \prime}+ \ddot{R}f^{\prime \prime} }{\kappa} - \frac{Rf^{\prime}-f}{2\kappa}\,,
\label{peff}
\end{eqnarray}
which are like effective energy-density and pressure due to the
curvature effects in contrast to the conventional $\rho$ and $P$ in
Eq.~(\ref{tmunu}) which are the energy-density and pressure of the
fluid present in the FRW spacetime. In this article we assume the
fluid to be barotropic so that its equation of state is
\begin{eqnarray}
P=\omega \rho\,,
\label{eqns}
\end{eqnarray}
where $\omega$ is zero for dust and one-third for radiation. It must
be noted that $u_\mu$ in Eq.~(\ref{tmunu}) is the
four-velocity of a fluid element and $u_\mu u^\mu = -1\,.$
%%%%%%%%%%%%%%%%%%%%%%%%%%%%%%%%%%%%%%%%%%%%%
\subsection{General bouncing conditions}
\label{bcond}

In this section we will primarily focus on the bouncing conditions in
the early universe keeping an eye on the energy conditions which are
followed by the matter sector. Some important results regarding the
bouncing conditions in FRW cosmologies were reported in
Ref.~\cite{Carloni:2005ii} where the authors very briefly inferred
about the possibility of cosmological bounces in flat FRW
cosmologies. In this article we will solely focus on cosmological
bounces in flat FRW models and discuss various interesting phenomena
related to these bounces.  Writing the time coordinate during the
bounce as $t=t_b$ and the Hubble parameter at bounce as $H_b =
H(t_b)$, the conditions for cosmological bounce are
\begin{eqnarray}  
H_b=0\,,\,\,\,\, {\rm and}\,\,\,\, \dot{H}_b>0\,.
\label{bconds}
\end{eqnarray}
Henceforth the subscript $b$ on any variable (which varies with time)
will specify its value at the time of bounce. In this convention one
can write the vanishing condition of the Hubble parameter during the bounce as
\begin{eqnarray}
3\frac{k}{a^{2}_{b}}=\frac{\kappa}{f^{\prime}_{b}}\left[ \rho_{b} + 
\frac{R_{b} f^{\prime}_{b}-f_{b}}{2\kappa} \right]\,.
\label{hzero}
\end{eqnarray}
The positivity of the time derivative of $H$ during bounce becomes 
$$2 \dot{H}_{b} = - \frac{\kappa}{f^{\prime}_{b}}\left[P_{b} +
  \frac{\dot{R}^{2}_{b}f^{\prime \prime
      \prime}_{b}+\ddot{R}_{b}f^{\prime \prime}_{b}}{\kappa} -
  \frac{R_{b} f^{\prime}_{b}-f_{b}}{2\kappa} \right]-
\frac{k}{a^{2}_{b}}>0\,,$$
which can also be written as
\begin{eqnarray}
\kappa \left( P_{b} + \rho _{b} \right) + \dot{R}^{2}_{b}f^{\prime
  \prime \prime}_{b}+\ddot{R}_{b}f^{\prime \prime}_{b} -
2 \frac{k f^{\prime}_{b}}{a^{2}_{b}}<0\,.
\label{hdotp}
\end{eqnarray}
If one assumes the hydrodynamic matter in the cosmological
background satisfies the standard weak energy condition as applied in GR: 
\begin{eqnarray}
\rho \ge 0\,,\,\,\,\,\rho + P \ge 0\,,
\label{econd}
\end{eqnarray}
then Eq.~(\ref{hdotp}) implies
\begin{equation}
\dot{R}^{2}_{b}f^{\prime \prime \prime}_{b}+\ddot{R}_{b}f^{\prime
  \prime}_{b} -2 \frac{k f^{\prime}_{b}}{a^{2}_{b}}<0\,.
\label{hdotp1}
\end{equation}
Using the form of $f(R)$ as given in Eq.~(\ref{ngrav}), 
the above equations predict that during a bounce the following
conditions must be fulfilled:
\begin{eqnarray}
\rho _{b} + \frac{(n-1)\alpha  R^{n}_{b}}{2\kappa} &=& 0 \,,
\label{eq:sign_alpha} 
\end{eqnarray}
for the spatially flat case, $k=0$.  When spatial curvature is zero,
$R_b = 6 \dot{H}_b$, and in such a case one can have a bouncing
universe in the quadratic gravity model, as specified in
Eq.~(\ref{ngrav}), when $\rho_b \ne 0$. Without matter there is no
bounce with the form of $f(R)$ chosen in Eq.~(\ref{ngrav}) in the
spatially flat FRW model.  More over as $n >1$ it can be seen from
Eq.~\eqref{eq:sign_alpha} that $\alpha < 0$.

For the sake of completeness we also briefly state that there can be
cosmological models in $f(R)$ theory where the presence of matter is
not absolutely important for the bouncing phenomenon. If one takes 
\begin{eqnarray}
f(R)= R + \beta R^2 + \gamma R^3\,,
\label{polinfr}
\end{eqnarray}
where $\beta$ and $\gamma$ are real numbers it can be shown that this
form of $f(R)$ is capable of producing cosmological bounce without
matter. In particular for the flat FRW spacetime in absence of any
matter the bouncing condition becomes
\begin{eqnarray}
(R_bf'_b-f_b)=0\,,
\label{bcondwm}
\end{eqnarray}
predicting that 
\begin{eqnarray}
R_b=-\frac{\beta}{2\gamma}\,.
\label{rbn}
\end{eqnarray}
As $R_b$ is positive definite for matter-less bounce in the flat FRW
universe, one has to assume that $\gamma$ and $\beta$ must have
different signs. In the present case if one demands that $f'(R)>0$,
such that the effective gravitational constant remains positive as
discussed in Ref.~\cite{Barrow:1988xh}, the coefficients $\beta$ and
$\gamma$ must satisfy the following inequality
\begin{eqnarray}
0 < \beta^2 \le 3\gamma\,.
\label{bgama}
\end{eqnarray}
From Eq.~(\ref{rbn}) it was seen that $\beta$ and $\gamma$ should have
opposite sign whereas from the above equation it can be uniquely said
that for a bouncing solution $\gamma > 0$ and $\beta < 0$.  With the
form of $f(R)$ as given in Eq.~(\ref{polinfr}) one can easily verify
that the third order gravitational action arising from it can also
support a cosmological bounce in flat FRW universe in the presence of
matter.

The above discussion on the possibility of bounces, in various $f(R)$
theories, was based on fact that the Hubble parameter must vanish at
$t=t_b$ as stated in Eq.~(\ref{hzero}). We did not discuss about the
second condition, the positivity of the rate of change of the Hubble
parameter at $t=t_b$, as given in Eq.~(\ref{hdotp1}). The main reason
for doing so is that these two conditions are independent. For a
spatially flat universe, the first condition of bounce in
Eq.~(\ref{hzero}) only depends on the value of $R_b$ \footnote{In this
  case $\rho_b$ is fixed when one specifies $R_b$.} whereas the second
condition in Eq.~(\ref{hdotp1}) depends only on $R_b$, $\dot{R}_b$ and
$\ddot{R}_b$. But all of these parameters are not independent of each
other. It can be easily shown that the system of equations governing
the dynamics of a flat FRW spacetime around a cosmological bounce, in
presence of barotropic matter, can be determined if one knows the
values of just two independent parameters $R_b$ and
$\dot{R}_b$. Whereas for a similar bounce in absence of matter the
value of $R_b$ is a constant as in Eq.~(\ref{rbn}), and the only
independent parameter left is that of $\dot{R}_b$. The above
discussion shows that the first bouncing condition, in
Eq.~(\ref{hzero}), only depends upon the value of $R_b$ and the second
bouncing condition, in Eq.~(\ref{hdotp1}), essentially only depends
upon the values of $R_b$ and $\dot{R}_b$ \footnote{As $\ddot{R}_b$ is
  is fixed once the values of $R_b$ and $\dot{R}_b$ are
  known.}. Consequently, if in a cosmological model the first bouncing
condition is satisfied for some value of $R_b$ the second bouncing
condition can also be satisfied by suitably choosing some value of
$\dot{R}_b$.

We have seen that some form of $f(R)$ can only produce a cosmological
bounce in presence of matter and some other form of $f(R)$ is capable
of producing a cosmological bounce in absence of matter. A natural
question arises at this point, whether there can be any quantitative
reasoning based on which one can differentiate these two forms of
$f(R)$.  We partially answer this in the next section where we present
a brief analysis about the form of $f(R)$ which can give rise to a
cosmological bounce, in absence of matter, in a flat FRW universe. The
answer is partial because the form of $f(R)$ which can produce
cosmological bounce in absence of matter may also be capable of
producing a cosmological bounce in the presence of matter.
%%%%%%%%%%%%%%%%%%
\subsection{Conditions for matter-less cosmological bounce in the flat FRW 
universe}
\label{mlb} 
In the most general case one can state the following facts about a
bouncing universe. {\it Given an} $f(R)$ {\it such that} $(Rf'-f)$
{\it has at least a positive root, then there always exists a
  matter-less bounce in the spatially flat} FRW {\it metric. Further,
  if} $\lim_{R \to 0} f(R) = 0$, {\it then the above condition implies
  that} $f'''(R)$ {\it cannot be identically zero for a matter-less
  bounce.} The proof of the above statements follow. In a matter-less
universe for $k=0$, if the Hubble parameter vanishes at any particular
value of the Ricci scalar then one must have
\begin{eqnarray}
Rf'-f=0\,.
\label{prf1}
\end{eqnarray}
More over in our case for $k=0$, the value of the Ricci scalar when
the Hubble parameter vanishes is given by  
\begin{equation}
R=6\dot{H}\,.
\label{rbc}
\end{equation}  
If for a particular $f(R)$ the function $Rf'-f$ has at least one
positive root $R=R_b$ then it implies $H=0$ at $R=R_b$. More over from
the above equation one can see that $\dot{H}$ remains positive when
$R=R_b$. The last two conditions, on the Hubble parameter and its time
derivative, specifies that there is a cosmological bounce at $R=R_b$.

If one also assumes that $\lim_{R \to 0} f(R) = 0$ (eliminating any
$f(R)$ with a cosmological constant like
term in it) then one can write
\begin{equation}
Rf'-f=R(R-R_{b})g(R)
\end{equation}
where $g(R)$ is nonsingular at $R=0$ and $R=R_{b}$.  Differentiating
the last equation one gets
\begin{equation}
Rf''=(2R-R_{b})g+R(R-R_{b})g'\,.
\end{equation}
Differentiating again, and using the above equation, one gets
\begin{equation}
R^2f'''=R_{b}g+R(3R-R_{b})g'+R^{2}(R-R_{b})g''\,.
\end{equation}
Now we make the proposition that a necessary condition for matter-less
bounce is that $ R^{2}f''' $ is not identically zero for the
particular $f(R)$. To prove this, we assume the contrary, i.e. 
$R^{2}f''' $ is identically zero for the particular form of $f(R)$. In
the present case then the function $ g(R) $ must be the solution of
the differential equation
\begin{equation}
R^{2}(R-R_{b})g'' + R(3R-R_{b})g' + R_{b} g = 0\,.
\end{equation}
This equation has an exact solution of the form
\begin{equation}
g(R)=\frac{C_{1}}{R(R-R_{b})}+\frac{C_{2}R}{2(R-R_{b})}\,,
\end{equation}
where $C_1$ and $C_2$ are arbitrary integration constants.  The
solution diverges at $R=R_{b}$ for all arbitrary constant values of
$C_1$ and $C_2$. This implies that in the most general case $g(R)$ is
singular at $R=R_{b}$, defying our assumption that $g(R)$ is non
singular at $R_b$. Consequently the assumption that $R^{2}f'''$ is
identically zero cannot be true. The last statement also implies that
$f'''$ cannot be identically zero in an interval of $R$ which contains
the bouncing phase.

Before ending this section we must remind the reader that the
possibility of a matter-less bounce in the flat FRW universe, in metric
$f(R)$ gravity, does not exclude the possibility of a cosmological
bounce in presence of  matter satisfying the energy conditions as given in
Eq.~(\ref{econd}). On the other hand there can be situations where one
cannot have a cosmological bounce in HD gravity in the
absence of matter as discussed in the case of quadratic
gravity.
%%%%%%%%%%%%%%%%%%%%%%%%%%%%%%%%%%%%%%%%%%%%%%%%%%%%%%%%%%%%%%%%%%%%%%%%%%%%
\subsection{Stability analysis in quadratic gravity with negative}
\label{stab}

The form of $f(R)$ in Eq.~(\ref{ngrav}) for $\alpha<0$ suggests that
$f'(R)$ may not be positive for all values of $R$. Particularly, if
$n=2$ then $f'(R)=1+2\alpha R$ which implies that for $R > \Lambda =
1/(2|\alpha|)$, $f'(R) < 0$ making the theory unstable as the
effective gravitational constant becomes negative.  During the
bouncing period the value of $R$ remains below the instability scale,
$\Lambda$, as can be verified from the plot of the logarithm of
$f'(R)$ (which is proportional to the scalar field $\varphi_0$ in the
Einstein frame, introduced in a later section), in Fig.~(\ref{phi1}). The
instability in the theory can show up near the end stages of the
bouncing period where $R$ can reach the instability scale.  This issue
can be addressed in broadly two different ways. 
\begin{enumerate}

\item One may choose the value of $\alpha$ in a particular way so that the
  instability  scale $\Lambda$ remains less than the value of the
  Ricci scalar at the present epoch of cosmic expansion. In such a
  case $f'(R)>0$ throughout the cosmic evolution if the initial value
  of the Ricci scalar during the contracting phase is less than
  $\Lambda$. This method of evading the instability may be practical
  but not consistent as the instability may show up in the future
  unless the theory of gravity itself changes.
   
\item If $\Lambda$ has a high value then the form of quadratic $f(R)$
  can certainly become unstable much earlier in our cosmic history. In
  this case if one wants to study the cosmological bounce in the
  quadratic theory then one has to stabilize the theory.  To make the
  theory stable one can enlarge the form of $f(R)$, by adding suitable
  terms to it, in such a way that the new terms in $f(R)$ makes the
  theory stable and at the same time does not change the nature of the
  cosmological bounce as predicted from the simple form of quadratic
  gravity as given in Eq.~(\ref{ngrav}). Consequently one may work
  with the quadratic form of $f(R)$ to investigate the time period
  near the cosmological bounce. To execute the above plan one may 
  generalize the form of $f(R)$ as:
\begin{eqnarray}
f(R)= R + \alpha R^2 + G(R)\,,
\label{genfr}
\end{eqnarray}
where $G(R)$ is a function of $R$ which is negligible during the
bouncing period.
A probable form of $G(R)$ can be
\begin{eqnarray}  
G(R)= \sum_n \xi_n (R-R_c)^{n}\,,
\label{grdef}
\end{eqnarray}
where $\xi_n$s are constants, $R_c$ is another constant whose
dimension is that of the Ricci scalar and $n (\ne 0)$ is an integer
which can take both positive and negative values. 

There can be a multiple number of ways in which one can choose a
$G(R)$. In this article we give a simple form of $G(R)$ which can
reasonably stabilize the quadratic HD theory.  If one sets $n=3$,
$R_c=0$ and $\xi_3>0$ and all other $\xi_n=0$ for $n\ne 3$, then
$f(R)$ becomes a cubic polynomial of $R$ and in such a case if $3 \xi_3 >
\alpha^2$ one can assure that $f'(R)>0$ for all values of $R$, as
shown in Eq.~(\ref{bgama}). By choosing suitable values of $|\alpha|$
one can easily satisfy the condition
$$G(R) \equiv \xi_3 R^3 < R + \alpha R^2\,,$$ if $R$ remains lower
than $\Lambda$. When $R\to \Lambda$ the cubic term starts to
contribute in the expression of $f(R)$ and in this limit the
energy-density of the hydrodynamic matter, which supports the
cosmological bounce tends to vanish due to cosmic expansion. In
general near the bounce the above form of $G(R)$ is two orders of
magnitude less than the pure quadratic term in the enlarged expression
of $f(R)$ if $\alpha \sim 10^{11-12}\,M_P^{-2}$ and $R<\Lambda$. Here
$M_P\sim 10^{19}\,{\rm GeV}$ is the Planck mass.  Consequently, in
this case the cubic term can be neglected during the brief bouncing
phase and one may treat the cosmological bounce problem in the
quadratic gravity paradigm.
\end{enumerate}

The above discussion on the stabilization of the theory of gravity
near the bouncing point shows that the HD quadratic gravity theory is
more like an effective theory of gravity. Near the bouncing point only
the terms up to quadratic order in the Ricci scalar contributes in the
gravitational action whereas away from the bounce other terms in the
action (of the general theory of gravity) becomes active. As the effect
of the function $G(R)$ can be neglected during cosmological bounce
predicted from the quadratic $f(R)$ in presence of matter we will
henceforth not discuss the effects of $G(R)$ in the bouncing
phenomenon. In this article we do not claim to give a full
cosmological evolution of the early universe to the present
universe. The main aim of this article is to describe the bouncing
phase using a quadratic gravity theory where one may require the
$G(R)$ function to avoid the intrinsic instability related to the
quadratic HD theory.

\begin{figure}[t!]
\centering
\includegraphics[scale=.8]{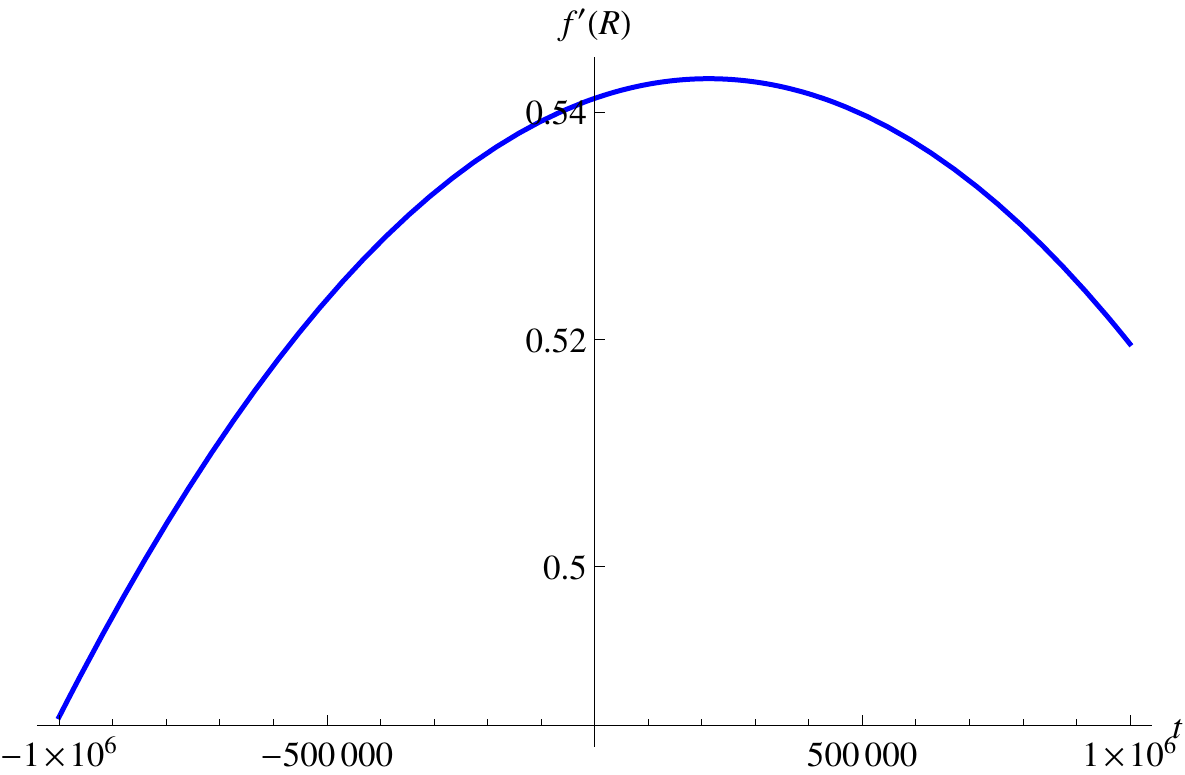}
\caption{Plot of $F(R)=f'(R)$ with respect to time in the Jordan frame
  in a radiation induced bounce in the quadratic gravity model. Here
  Ricci scalar and cosmic time are expressed in Planck units. The details of the
  model is presented in subsection \ref{einf} where the topic of radiation induced
  bounce is presented.}
\label{fprime}
\end{figure}
%%%%%%%%%%%%%%%%%%%%%%% 
In Fig.~\ref{fprime} we present a specimen plot of $f'(R)$ with
respect to time, during the radiation induced bounce in the quadratic
gravity model (disregarding the effects of $G(R)$) and show that
$f'(R)$ remains positive during the bouncing regime. If $G(R)$ is
included then the nature of the curve will change during the end phases
of the bouncing period.  The particulars of the model which produces
the features of the plot will be explained in section \ref{cp} where
we discuss the detailed analysis of the bouncing mechanism.

%%%%%%%%%%%%%%%%%%%%%%%%%%%%%%%%%%%%%%%%%%%%%
\section{Relativity of conformally connected frames}
\label{cp}
In this section we will present a detailed analysis of the bounce
mechanism in HD gravity. As HD gravity theories in the presence of
matter is a difficult theory to solve analytically we will take
recourse to a method which relies on {\it relativity of conformally
  connected frames}. The HD theory is defined in a frame which is
often called the matter frame \cite{Carroll:2003wy} or the Jordan
frame. As one can always cast a metric $f(R)$ theory of gravity in the
Jordan frame, where the theory of gravity looks like a Brans-Dicke
theory with the Brans-Dicke parameter $\omega_0=0$
\cite{Sotiriou:2008rp} without doing any conformal transformation, we
will mostly call the conformal frame of the HD theory to be the Jordan
frame.

Once the HD gravity theory is set in the matter or Jordan frame one
can make a conformal transformation from the Jordan frame to the
Einstein frame where the corresponding theory will be described by the
Einstein's equation in GR and the dynamics of a scalar field which is
minimally coupled to gravity and non-minimally coupled to matter.  It
is a general belief that the description of gravitational physics in
the Jordan frame is equivalent to that in the Einstein frame. In a
previous work in Ref.~\cite{Faraoni:1999hp} the authors, who have
studied the nature of gravitational waves in both conformal frames,
have shown that this belief need not be always true. In
Ref.~\cite{Capozziello:2006dj} the authors specifically mentioned that
the cosmological evolution in Jordan frame and the Einstein frame can
be different in $f(R)$ theories. In this article
we will observe a clear case where the description of cosmology in
these two conformal frames are different. If there is a cosmological
bounce in the flat FRW universe in the Jordan frame there cannot be
any analogous bounce in the Einstein frame.

As the two conformal frames in our case do not predict equivalent
bouncing cosmologies, we have to choose the frame which is physical,
as discussed in \cite{Capozziello:2006dj}. In our specific case as the
HD theory is initially presented in the Jordan frame we will take the
Jordan frame to be the physical frame. The Einstein frame serves as an
auxiliary frame in which the calculations are done for
simplicity. This distinction between the physical frame and the
auxiliary frame was previously discussed in the section regarding
conformal transformations in Ref.~\cite{ DeFelice:2010aj}.  The
conformal transformations from the matter frame to the Einstein frame
have been studied earlier \cite{Whitt:1984pd, Barrow:1988xh,
  DeFelice:2010aj, Wands:1993uu}.
%%%%%%%%%%%%%%%%%%%%%%%%%%%%%%%%%%%%%%%%%%%%%%%%%%%%%%%%%%%%%%%%%%%
\subsection{Einstein frame description of 
  higher-derivative gravity}
\begin{figure}[t!]
\centering
\includegraphics[scale=1]{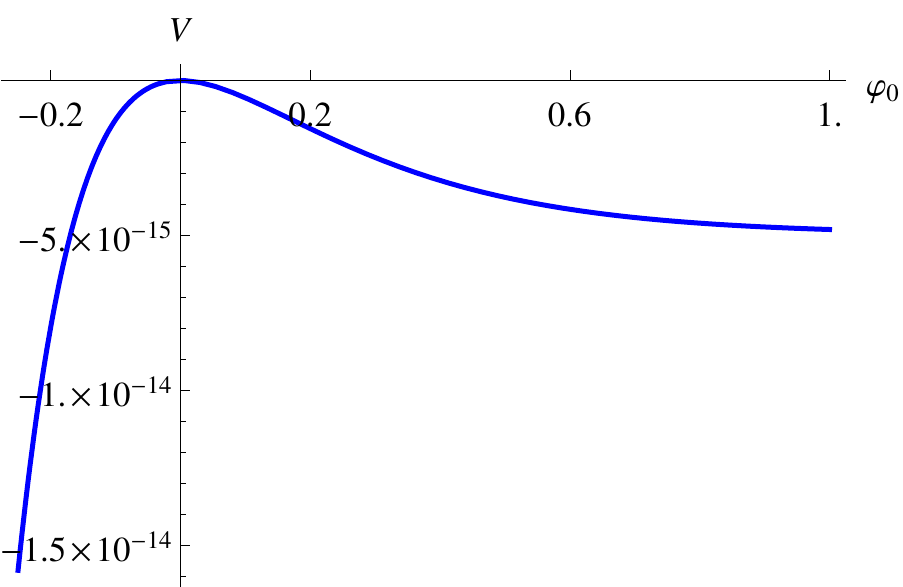}
\caption{Figure showing the nature of the scalar field potential in
  the Einstein frame where $f(R)=R+\alpha R^2$. In the above figure
  the field and its potential both are expressed in Planck units
  (where the Planck mass is set as unity) in which
  $\alpha=-10^{12}$. The reason for such a magnitude of $\alpha$ is
  explained in the text below.  }
\label{phi_pot}
\end{figure}

Starting from the dynamical equation of
motion of the gravitational field in the higher-derivative $f(R)$
theory as given in Eq.~(\ref{ein1}) one can apply a conformal
transformation as
\begin{eqnarray}
\tilde{g}_{\mu \nu} = F(R)g_{\mu \nu}\,,
\label{conf}
\end{eqnarray}
where $$F(R)\equiv \frac{d f(R)}{dR}\,,$$ and obtain an effective
equation of gravitational dynamics in the Einstein frame as
\begin{eqnarray}
\tilde{R}^\mu_\nu - \frac12 \delta^\mu_\nu \tilde{R}
= \kappa  \tilde{S}^\mu_\nu\,.
\label{neqn}
\end{eqnarray}
The energy-momentum tensor is given as
\begin{eqnarray}
\tilde{S}^\mu_\nu =\tilde{T}^\mu_\nu + \tilde{\mathcal T}^\mu_\nu\,,
\label{mttilde}
\end{eqnarray}
where $\tilde{T}^\mu_\nu$ is the energy-momentum tensor of a real
scalar field $\varphi_0$ defined as
\begin{eqnarray}
\varphi_0 \equiv \sqrt{\frac{3}{2\kappa}}\ln F\,.
\label{phidef}
\end{eqnarray}
Here the subscript $0$ in $\varphi_0$ specifies that the scalar field
controls the background cosmological evolution in the Einstein
frame. Later we will deal with perturbations on this background,
$\delta \varphi$, to study the evolution of the cosmological
perturbations in a bouncing universe.  For the particular model of
quadratic $f(R)$ one can always satisfy the condition $F>0$ near the
bounce point and so the above definition of the scalar field will
pose no problem. The potential of the scalar field in the Einstein
frame is given by
\begin{eqnarray}
V(\varphi_0)=\frac{R F-f}{2\kappa F^2}\,.
\label{vexp}
\end{eqnarray}
The $\varphi_0$ dependence of $V$ is obtained by expressing $R$ as a
function of $\varphi_0$ using Eq.~(\ref{phidef}) and the definition of
$F(R)$. For an example, if one assumes $n=2$ in Eq.~(\ref{ngrav}),
then the expression of $V(\varphi_0)$ is of the following form:
\begin{eqnarray}
V(\varphi_0)=\frac{1}{8\kappa \alpha}\left(1-e^{-\sqrt{2\kappa/3}\,\,\varphi_0}\right)^2\,,
\label{vphi}
\end{eqnarray}
whose plot is shown in Fig.~\ref{phi_pot}. 
The second term in $\tilde{S}^\mu_\nu$ is $\tilde{\mathcal T}^\mu_\nu$ and it 
is related to the actual energy-momentum tensor of the fluid as given
in Eq.~(\ref{tmunu}) via
\begin{eqnarray}
\tilde{\mathcal T}^\mu_\nu = \frac{T^\mu_\nu}{F^2}\,. 
\label{conft}
\end{eqnarray}
Using the transformed metric $\tilde{g}_{\alpha \beta}$ to lower the
contravariant index in $\tilde{\mathcal T}^\mu_\nu$ one can rewrite the
above equation in the following form
\begin{eqnarray}
\tilde{\mathcal T}_{\mu\nu} = (\tilde{\rho} + \tilde{P})\tilde{u}_\mu 
\tilde{u}_\nu + \tilde{P} \tilde{g}_{\mu\nu}\,,
\label{tmunup}
\end{eqnarray}
where the new energy-density, pressure and the 4-velocity of the
hydrodynamic fluid element are
\begin{eqnarray}
\tilde{\rho}=\frac{\rho}{F^2}\,,\,\,\tilde{P}=\frac{P}{F^2}\,,\,\,
\tilde{u}_\mu \equiv \sqrt{F} u_\mu\,.
\label{nrhop} 
\end{eqnarray}
The new 4-vectors $\tilde{u}_\mu$ are defined in such a way that
$\tilde{g}^{\mu \nu}\tilde{u}_\mu \tilde{u}_\nu=-1$.  It is
interesting to note that if the equation of state of the fluid in the
higher-derivative description was $P=\omega \rho$ then in the
Einstein frame the equation of state of the fluid remains the same.

One can now interpret Eq.~(\ref{neqn}) as the equivalent FRW
description, in GR, of the HD gravitational theory
where the new line element written in  terms of the
conformally transformed metric $\tilde{g}_{\mu\nu}$ looks like
\begin{eqnarray}
d\tilde{s}^2 = -d\tilde{t}^2 + \tilde{a}^2(\tilde{t})
\left[\frac{dr^2}{1-kr^2} + r^2(d\theta^2 + \sin^2 \theta\,\,
d\varphi_0^2)\right]\,,
\label{nfrw}
\end{eqnarray}
where $$d\tilde{t}=\sqrt{F(R)}\,\,dt\,,\,\,\,\,\tilde{a}(t)=
\sqrt{F(R)}\,\,a(t)\,.$$ The above equations provide us the redefined
time in the Einstein frame corresponding to the time
variable appearing in the original FRW metric in Eq.~(\ref{frw}) in
the higher-derivative theory, as
\begin{eqnarray}
\tilde{t}= \int_{t_b}^t \sqrt{F(R)}\,\,dt\,,
\label{newt}
\end{eqnarray}
where the lower integration limit is set at the bouncing time in the
Jordan frame. The relation between $t$ and $\tilde{t}$ shows that
whatever be the value of $t_{b}$ , the time corresponding to it in the
Einstein frame, $\tilde{t} = 0 $.

  The HD effects can be important in the early universe when the
  Hubble parameter was big, $H\sim 10^{12-13}\,\rm{GeV}$. For this
  value of the Hubble parameter one obtains $R \sim
  10^{26}\,\rm{GeV}^{2}$ assuming a very slow change of $H$ with
  time. If the HD correction $\alpha R^2$ becomes effective near about
  this time then $|\alpha| \sim 1/R$ yielding $|\alpha| \sim
  10^{-26}\,{\rm GeV}^{-2} = 10^{12}\,M^{-2}_P$.  The relationship of
  quadratic gravity, giving rise to a matter induced cosmological
  bounce, and a more general cubic form of $f(R)$ containing more
  ingredients in it was discussed in subsection \ref{stab}. The chosen
  value of $\alpha$ in our case is such that one may use the quadratic
  form of HD gravity theory near the bounce.

The relation between the Hubble parameter, $H$, appearing in the
higher-derivative gravity theory and the effective Einstein frame
Hubble parameter $\tilde{H}$, defined as $\tilde{a}'/\tilde{a}$, is
\begin{eqnarray}
H=\sqrt{F}\left(\tilde{H}-\sqrt{\frac{\kappa}{6}}\,\varphi_0'\right)\,,
\label{hhtilde}
\end{eqnarray}
where the prime now stands for $d/d\tilde{t}$ \footnote{At this point
  we must alert the reader that we will use a superscript prime to
  symbolize three different quantities in this article. In subsection
  \ref{bcond} the prime stood for a derivative with respect to the
  Ricci scalar. In this section the prime stands for a derivative with
  respect to the time variable in the Einstein frame. In the section
  \ref{perts} the prime will stand for a derivative with respect to
  the conformal time.}. The Einstein frame Hubble parameter satisfies
the following Friedman equation
\begin{eqnarray}
\tilde{H}^2 + \frac{k}{\tilde{a}^2}= \frac{\kappa}{3}(\rho_{\varphi_0} +
\tilde{\rho})\,,
\label{fridm1}
\end{eqnarray}
where
\begin{eqnarray}
\rho_{\varphi_0} = \frac12 \varphi_0'^2
+ V(\varphi_0)\,,
\label{phieng}
\end{eqnarray}
and the expression for $\tilde{\rho}$ is given in Eq.~(\ref{nrhop}). The
other relevant equation for our case is
\begin{eqnarray} 
\tilde{H}'= \frac{k}{\tilde{a}^2}-\frac{\kappa}{2}[\varphi_0'^2 +
(1+\omega)\tilde{\rho}]\,,
\label{hdot}
\end{eqnarray}
where $\omega$ appears in the equation of state of the fluid whose
energy-momentum tensor is given in Eq.~(\ref{tmunup}).  Taking the
trace of Einstein equation one gets the time evolution equation for
the scalar field $\varphi_0$ as:
\begin{eqnarray}
\varphi_0''+ 3\tilde{H} \varphi_0' +
\frac{dV}{d\varphi_0}=\sqrt{\frac{\kappa}{6}}(1-3\omega)\tilde{\rho}\,.
\label{phiev}
\end{eqnarray}
The above equation shows that time development of the scalar field
depends upon the presence of the fluid energy-density. There is a
coupling of matter and the scalar field in the Einstein frame. Only in
the case of radiation does the right hand side of the above equation
becomes zero and the scalar field can evolve in the conventional
way. A radiation bath does not affect the scalar field dynamics due to
conformal symmetry. The evolution of the energy density in the
Einstein frame follows:
\begin{eqnarray}
\tilde{\rho}' + \sqrt{\frac{\kappa}{6}}(1-3\omega)\tilde{\rho}\varphi_0'
+3\tilde{H}\tilde{\rho}(1+\omega)=0\,. 
\label{rtildev}
\end{eqnarray}
The set set of equations, Eq.~(\ref{fridm1}), Eq.~(\ref{hdot}),
Eq.~(\ref{phiev}) and Eq.~(\ref{rtildev}), are the main equations
which dictate the behavior of the system in the Einstein frame.

From the dynamical equations of the cosmological model presented in
the Einstein frame it may seem that one must require the values of a
plethora of variables at the bounce time to run the system
analytically or numerically. But a closer inspection reveals that in
the case of the spatially flat FRW spacetime the values of just two
quantities, at the bouncing time, are enough to specify the past and
future of the universe around the bouncing point. In this article we
will take these two quantities to be $\varphi_0$ and $\varphi_0'$ at
$\tilde{t}=0$.

Next we state an interesting relationship of the bounce in the Jordan
frame and its analogous behavior in the Einstein frame.

%%%%%%%%%%%%%%%%%%%%%%%%%%%%%%%%%%%%%%
\subsection{On the absence of bounce in the Einstein frame}
\label{abef}
{\it A cosmological bounce in the flat} FRW {\it space in metric}
$f(R)$ {\it gravity theory in the Jordan frame will never have an
  analogous cosmological bounce in the conformally related Einstein
  frame if matter satisfies the condition} $\rho + P \ge 0$. {\it The
  statement holds even in the absence of matter.} The proof of the
above statements follows. If there is a cosmological bounce in the
Jordan frame for some particular form of $f(R)$ in the flat FRW
universe then at the time of bounce $t=t_b$ one must have $H_b=0$
corresponding to which one gets
$\tilde{H}(0)=\sqrt{\kappa/6}\,\varphi_0'(0)$ from Eq.~(\ref{hhtilde})
in the conformally related Einstein frame. The last relation shows
that $\tilde{H}(0)$ need not be zero at $\tilde{t}=0$ which
corresponds to the bouncing time in the Jordan frame. More over from
Eq.~(\ref{hdot}) one one can see that $\tilde{H}'(0)=
-\frac{\kappa}{2}(\varphi_0'^2 + \tilde{\rho} + \tilde{P})<0$ if
matter respects the energy conditions as stipulated in
Eq.~(\ref{econd}). Consequently the cosmological bounce in the
higher-derivative theory does not correspond to an analogous bounce in
the conformally related Einstein frame.

The above proof is valid only for the cases which involve cosmological
bounces in the flat FRW universe. If one takes the spatial curvature
of the FRW universe to be non-zero one may have bounces in both the
frames only when the spatial curvature is positive, $k=1$.  A set of
sufficient conditions predicting simultaneous bounces in both the
frames are presented in the following discussion. If the bouncing
conditions in the Einstein frame are satisfied then there will always
be a simultaneous bounce in the Jordan frame when
\begin{eqnarray}
\varphi_0^{\prime} = 0\,,\,\,\,\, \varphi_0^{\prime \prime} < 0\,
\label{sufc}
\end{eqnarray}
at $\tilde{t}=0$ \footnote{Here we have represented the conditions of
  a simultaneous bounce in the Einstein frame variables. If one wishes
  one may easily express these equations in terms of the Jordan frame
  variable $R$ and its time derivatives by using Eq.~(\ref{phidef})
  and the form of $F(R)$.}.  We see from Eq.~\eqref{hhtilde} that if
$\varphi_0^{\prime}(0) = 0$ then both $H$ and $\tilde{H}$ can be zero
simultaneously at $t=\tilde{t}=0$.  If $\varphi_0^{\prime}(0)$ is indeed
zero then Eq ~(\ref{hdot}) reduces to,
\begin{eqnarray} 
\tilde{H}'(0)=
\frac{k}{\tilde{a}^2(0)}-\frac{\kappa}{2}\left[\tilde{\rho}(0)+
\tilde{P}(0)\right]\,.
\end{eqnarray}
The above equation shows that one of the prerequisites for a bouncing
solution in the Einstein frame is $k>0$ such that
$\tilde{H}^{\prime}(0)$ can be positive.  Differentiating
Eq.~(\ref{hhtilde}) with respect to the cosmological time, $t$, we get
that at $t=\tilde{t}=0$,
\begin{eqnarray}
\dot{H} &=& F\left( \tilde{H}^{\prime} - 
\sqrt{\frac{\kappa}{6}}\varphi_0^{\prime \prime} \right)\,,
\end{eqnarray}
which implies that a sufficient condition for a simultaneous
cosmological bounce in both frames, such that both $\dot{H}$ and
$\tilde{H}'$ remain positive at $t=\tilde{t}=0$, is
$\varphi_0^{\prime\prime}(0) < 0$.

%%%%%%%%%%%%%%%%%%%%%%%%%%%%%%%%%%%%%%%%%%%%%%%%%%%%%%%%%%%%%%%%%%%%%%%%
\section{Analysis of the bounces in the Jordan frame}
\label{numb}

In this section we discuss the cosmological bouncing phenomena in
detail using the specific the quadratic gravity $f(R)$ model and
verify the claims made so far. To study the bouncing patterns in the
flat FRW cosmologies we employ numerical techniques as analytic
techniques do not yield simple results for most of the cases in HD
theories of gravity. Although, the numerical solutions of the bouncing
equations could be done in the Jordan frame itself but we chose to
solve the dynamics of the system in the Einstein frame for two
important reasons. The first reason is to show specifically the
different descriptions, of the same cosmological event, in the two
conformally connected frames. This is what we call the relativity of
conformally connected frames. The second reason is related to the
physical understanding of the bouncing phenomenon in the Jordan
frame. Purely a numerical study in the Jordan frame will not answer
the typical questions regarding the qualitative nature of the
asymmetric bounces. On the other hand it will be seen that the scalar
field potential in the Einstein frame and the conditions of
cosmological bounce, as translated in the Einstein frame, themselves
qualitatively convey the physical nature of the bounces in the
conformally connected Jordan frame.

Although we use the Einstein frame to solve our problem,
it must be said that the Einstein frame description of the bouncing
phenomenon always involves a scalar potential which is unbounded from
below. It does not produce a difficult situation in our case because
we will see that most of the dynamics of the bouncing system can be
effectively described by the behavior of the scalar field near the top
of the potential. The HD description of the
cosmological bounce does not remain effective once the scalar field
has slightly rolled down in the unstable direction. 
\begin{figure}[t!]
\begin{minipage}[b]{0.5\linewidth}
\centering
\includegraphics[scale=.8]{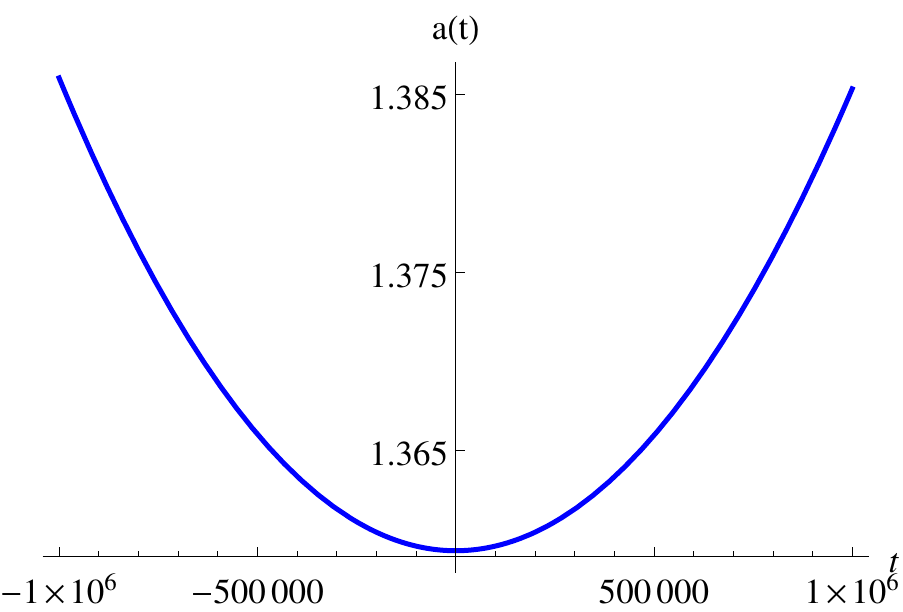}
\caption{Plot of the Cosmic Scale factor with time in the Jordan frame
 for quadratic gravity.} 
\label{a1}
\end{minipage}
\hspace{0.2cm}
\begin{minipage}[b]{0.5\linewidth}
\centering
\includegraphics[scale=.7]{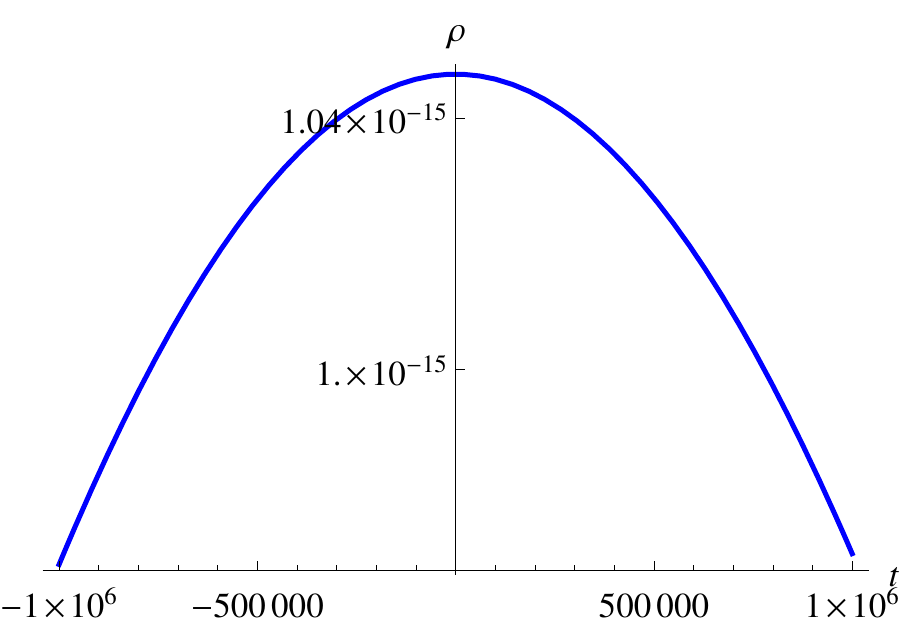}
\caption{Plot of matter density with time in the Jordan frame for 
quadratic gravity.}
\label{endens1}
\end{minipage}
\end{figure}

In the rest of this section we discuss the properties of cosmological
bounce in the presence of matter where $f(R)=R+\alpha R^2$.
Henceforth we will always use $\alpha= -10^{12}\,M_{P}^{-2}$, for
explicit numerical calculations. The reason for choosing this
numerical value of $\alpha$ was discussed in the last section.  The
most probable fluid, which could have been present, during the
cosmological bounce can be the perfect radiation fluid.  The radiation
energy density starts to build up during the contracting phase and
peaks during the bounce, after which it starts to get diluted and
practically vanishes at the end phase of expansion.  In the following
analysis we have assumed that bounce occurs in the Jordan frame at
$t=0$, which corresponds to $\tilde{t}=0$ in the Einstein frame. In
Planck units (where $M_{P}$ will be set as one), which we will use
henceforth, if one chooses $|\alpha| \sim 10^{12}$ then the time
period during which the HD gravity effects become important is given
by $-10^{6}\le t \le 10^{6}$ \footnote{In conventional units, this
  time period is roughly $-10^{-37}$s to $10^{-37}$s. Although we use
  the same time interval $-10^{6}\le t \le 10^{6}$ in both the
  Einstein frame and the Jordan frame actually the time-interval in
  the Einstein frame gets very slightly stretched in the Jordan frame
  due to the relation between $t$ and $\tilde{t}$. In this article we
  omit this stretching and work with the same time-interval in both
  the frames.}.  More over as we recreated the bounce in the Jordan
frame from its behavior in the Einstein frame, the relevant conditions
of bounce, as the energy density and other variables during bounce,
are specified in the Einstein frame.  In the Einstein frame one
requires only the values of two variables, $\varphi_0$ and
$\varphi_0'$, at the bouncing point to predict the cosmological
dynamics around the bounce.  Translated into the Jordan frame the
values of those two variables give us the Ricci scalar $R$ and its
time derivative $\dot{R}$ as
\begin{figure}[t!]
\centering
\includegraphics[scale=1.1]{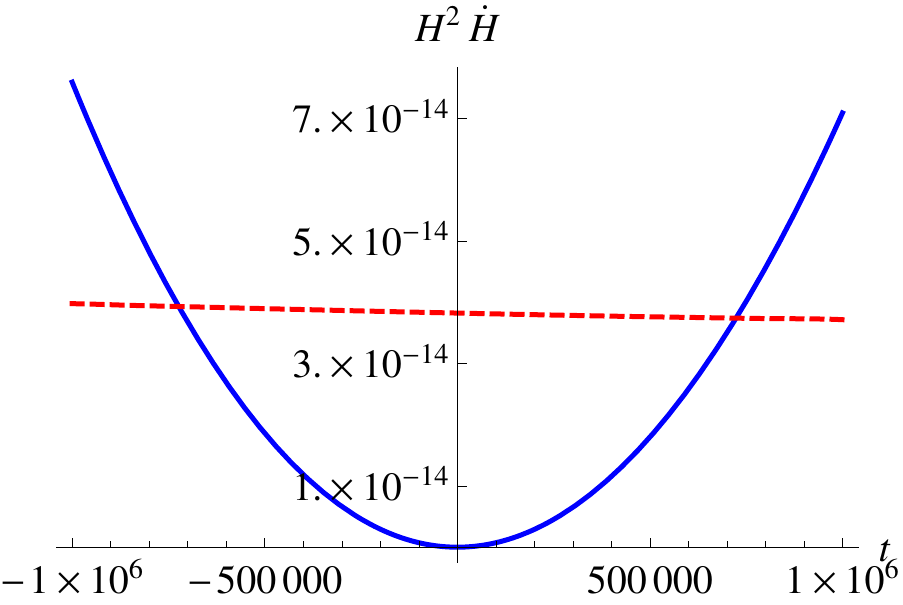}
\caption{Plot of the square of the Hubble parameter and its time
  derivative with respect to time in the Jordan frame for quadratic
  gravity model. The blue solid line represents $H^2 \times 50$ and
  the red dashed line represents ${\dot H}$.}
\label{hhdot1}
\end{figure}
%%%%%%%%%%%%%%%%%%%%%%%
\begin{eqnarray} 
R(0) = \frac{e^{\sqrt{2\kappa/3}\varphi_0(0)}-1}{2\alpha}\,,\,\,\,
\dot{R}(0) = \frac{\varphi_0'(0)}{2\alpha}\sqrt{\frac{2\kappa}{3}}
(1+2\alpha R(0))^{\frac{3}{2}}\,,
\end{eqnarray}
at the bouncing point. In the Jordan frame these two values specify
the cosmological state at the bouncing time.

The plot in, Fig.~\ref{a1}, shows the variation of the scale factor
during the cosmological bounce whereas Fig.~\ref{endens1} shows the
variation of the radiation energy density during the bounce. From
these figures it is apparent that the cosmological bounce in this case
is slightly asymmetric. The cause of this asymmetry is related to the
values of $\varphi_0$ and $\varphi_0'$ at $\tilde{t}=0$ in the Einstein
frame and will be explained in the next section. 

Fig.~\ref{hhdot1} shows, the variation of the square of the Hubble
parameter and the time derivative of the Hubble parameter, with respect
to time in the Jordan frame during the bouncing period. During the
bouncing period it is seen that the time derivative of the Hubble
parameter is practically a constant. During the start of the
contracting phase of the universe $\dot{H} < H^2$ and again during the
end of the expansion phase after bounce $\dot{H} < H^2$, indicating
that the initiation of the radiation dominated bouncing phase may be
approximated by a deflationary period and the end of the bouncing
phase may be approximated by an inflationary regime.
%%%%%%%%%%%%%%%%%%%%%%%%%%%%%%%%%%%%%%%%%%%%%%
\section{The corresponding analysis in the Einstein frame}
\label{einf}
\begin{figure}[t!]
\centering 
\includegraphics[scale=1] 
{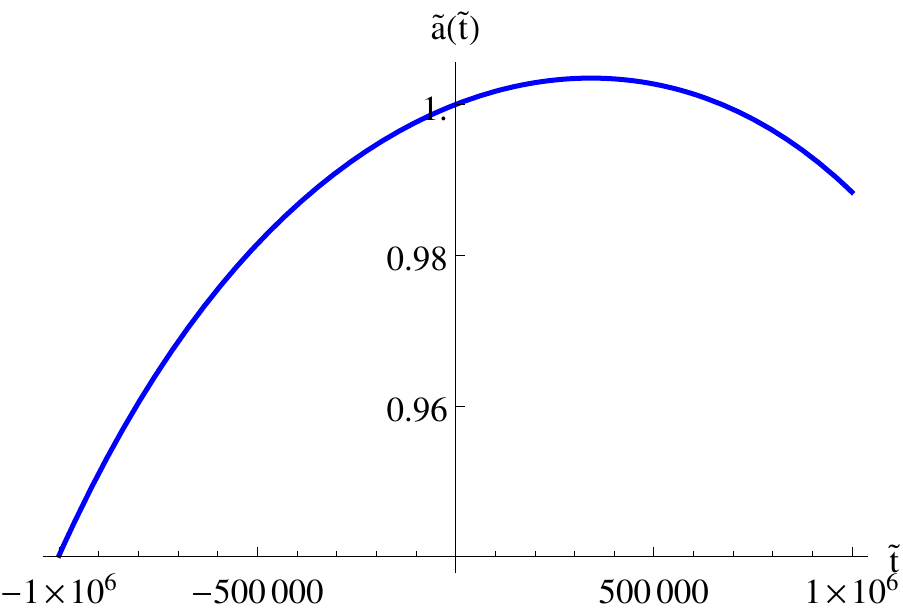}
\caption{The figure shows the plot of $\tilde{a}(t)$ with $\tilde{t}$ in
Einstein frame for the case where the bounce is induced by radiation.}
\label{ea1}
\end{figure}

In this section we specify the method by which we obtained the
bouncing solutions. The problem was to solve the system in the
Einstein frame and then using the methods in section \ref{cp}, convert
the solutions to the Jordan frame. It was stated before that for the
flat universe, one cannot have cosmological bounces in both the Jordan
and Einstein frames. This statement can be observed by the behavior
of the scale factors in the corresponding figures in the Jordan frame
and the Einstein frame. The figures, Fig.~\ref{a1} and Fig.~\ref{ea1}
show the differences for the case of quadratic gravity models where
bounce occurs in presence of a radiation fluid, in the Jordan
frame. In this section we specify the Einstein frame description of
the events which correspond to the cosmological bounces in the Jordan
frame.

In this section we mainly describe the Einstein frame behavior of
quadratic gravity, where $f(R)=R+\alpha R^2$.  From Fig.~\ref{ea1} it
is clear that there is no bounce in the Einstein frame. In the
Einstein frame the universe looks like an expanding phase
initially. The expansion gives way to a slow contraction during the
end of the stipulated time period in which the HD theory is
effective. From the plot of the potential of the scalar field
$\varphi_0$ given in Fig.~\ref{phi_pot} one can have an idea of the
behavior of this case in the Einstein frame. For positive values of
$\varphi_0$, the Ricci scalar in the Jordan frame turns out to be
negative (for a negative $\alpha$) ruling out any possibility of a
bounce.

\begin{figure}[t!]
\begin{minipage}[b]{0.5\linewidth}
\centering \includegraphics[scale=.55]{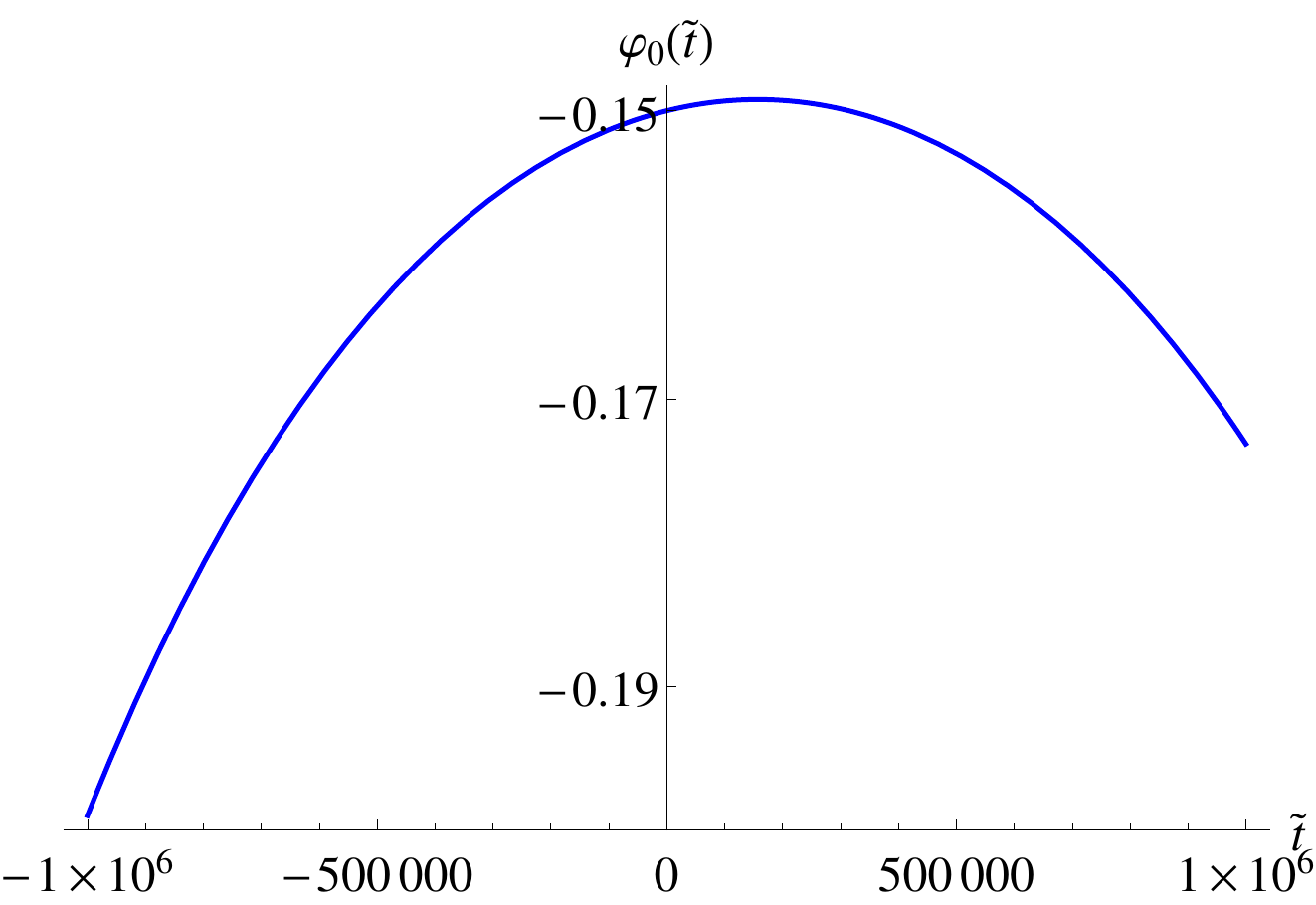}
\caption{Figure showing the plot of effective scalar 
field $\varphi_0$ with time in Einstein Frame in case of quadratic
$f(R)$. In the Einstein frame instead of an initial contraction there
is an initial expansion.}
\label{phi1}
\end{minipage}
\hspace{0.2cm}
\begin{minipage}[b]{0.5\linewidth}
\centering
\includegraphics[scale=.8]{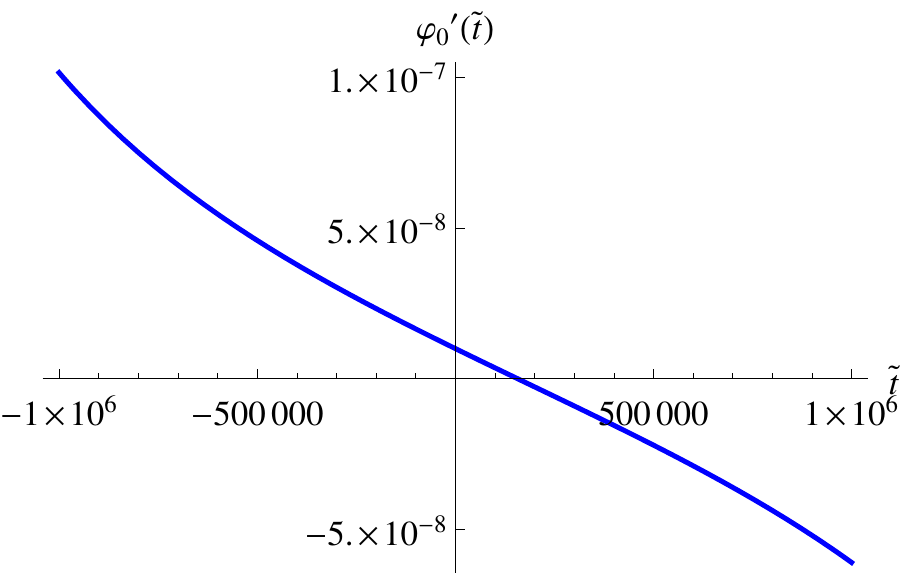}
\caption{Figure showing the plot of time derivative of 
effective scalar field $\varphi_0$ with time in Einstein Frame in case of 
quadratic $f(R)$. The figure shows the turning point of $\phi_0$ in
time where 
$\phi'=0$.}
\label{phip}
\end{minipage}
\end{figure}

The plot of the scalar field potential, Fig.~\ref{phi_pot}, shows that
the scalar field must remain sub-Planckian, $|\varphi_0|< M_{P}$,
through out the bouncing period.  The scalar field potential for
negative values of $\varphi_0$ is not bounded from below. For a
cosmological bounce, in the Jordan frame, the scalar field in the
Einstein frame must have some negative value at $\tilde{t}=0$ to start
with. If $\varphi_0'>0$ at $\tilde{t}=0$ then $\varphi_0$ increases in
time and tries to reach the top of the potential. In this case the
Hubble parameter in the Einstein frame remains positive for a long
time as the condition of bounce in the Jordan frame requires
$\tilde{H}>0$ at $\tilde{t}=0$ evident from Eq.~(\ref{phi0}).  In its
upward journey, the kinetic energy of the scalar field decreases as
shown by the decrement of $\varphi_0'$ in Fig.~\ref{phip}. This slows
down the development of the scalar field in its run towards the
turning point.  There is a point up to which $\varphi_0'$ remains
positive and then it flips sign and the scalar field rolls down the
steep side of the potential.  The bouncing condition,
Eq.~(\ref{phi0}), and the shape of the potential of the scalar field
produces the asymmetric growth of $\varphi_0$ in time and this
asymmetry is qualitatively reflected in the asymmetric bouncing nature
of the scale factor in the Jordan frame. If one wants a perfectly
symmetrical bounce in the Jordan frame then one has to implement the
condition $\varphi_0'=0$ at $\tilde{t}=0$ which makes the scalar
field's turning point at precisely $\tilde{t}=0$. One may also give a
negative value to $\varphi_0'$ at $\tilde{t}=0$ resulting in an
asymmetric bounce where the evolution of the scale factor, in the
Jordan frame, slows down in the contracting phase of the universe
unlike the one shown in Fig.~\ref{a1}, where $\varphi_0' > 0$ at
$\tilde{t}=0$. 

For the plots presented in this article, showing radiation dominated
bounce in the Jordan frame, we have chosen that at $\tilde{t}=0$ 
\begin{eqnarray}
\varphi_0 = -.15\,,\,\,\,\,
\varphi_0' = 10^{-8}\,.
\label{initphip} 
\end{eqnarray}
The smallness of $\varphi_0'$ and the position of
$\varphi_0$ in the scalar field potential dictates the
time scale of the bouncing phenomenon.  These two are the only
independent quantities, specifying whose values at the time of bounce,
dictates the dynamics of the cosmological system uniquely. 

\begin{figure}[t!]
\centering
\includegraphics[scale=1.1]{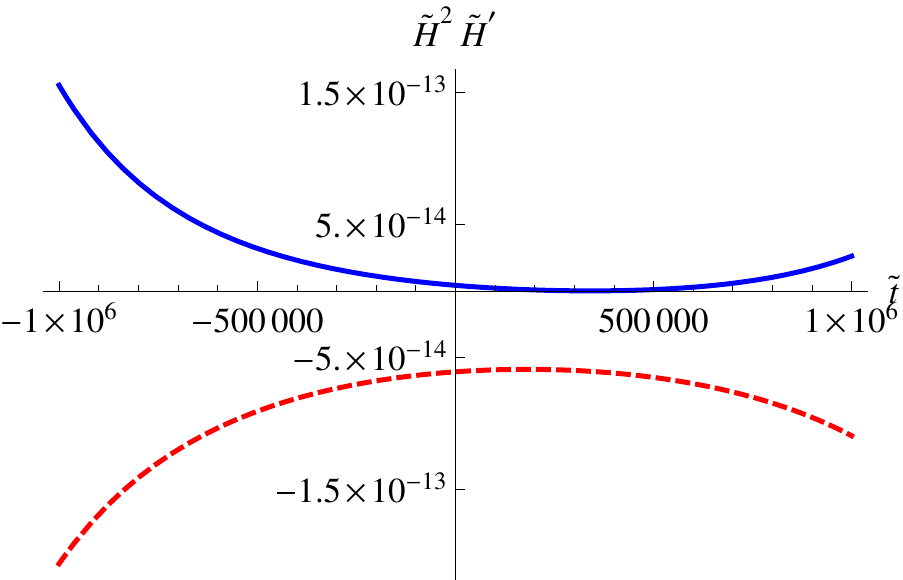}
\caption{The plot showing how $\tilde{H}^2$ and $\tilde{H}'$ vary with
  $\tilde{t}$ in the Einstein frame in case of quadratic $f(R)$. The blue
  solid line is $\tilde{H}^2\times 10$ and the red dashed line is
  $\tilde{H}'$.}
\label{hhdote1}
\end{figure}

The effective ambient hydrodynamic matter in the Einstein frame is
modelled by a fluid whose $\omega=1/3$, like conventional radiation,
but the energy density and pressure of this fluid depends upon the
scalar field strength. This coupling of the scalar field to the
original hydrodynamic fluid is referred as a non-minimal coupling of
the scalar field to matter. The state of hydrodynamic matter at
$\tilde{t}=0$ is obtained from the bouncing condition
in the Jordan frame $H(t=0)=0$, which implies 
\begin{eqnarray}
\tilde{H} &=& \sqrt{\frac{\kappa}{6}}\varphi_0^{\prime}\,,
\label{phi0}
\end{eqnarray} 
at $\tilde{t}=0$ in the Einstein frame, and Eq.~(\ref{fridm1}). The
aforementioned two conditions specifies 
\begin{eqnarray}
\tilde{\rho} &=& -V(\varphi_0)\,, 
\label{matter1}
\end{eqnarray}
at $\tilde{t}=0$, giving the energy content in the non-minimally
coupled sector.  Finally, the plot of the Hubble parameter and its
time derivative in the Einstein frame is shown in
Fig.~\ref{hhdote1}.
%%%%%%%%%%%%%%%%%%%%%%%%%%%%%%%%%%%%%%%%%%%%%%%%%%%%%%%%
\section{Scalar metric perturbations in the Einstein frame}
\label{perts}

The background bouncing FRW solutions, in the Jordan frame, can have
cosmological perturbations and these perturbations evolve in
spacetime. Cosmological perturbations in $f(R)$ gravity has been
studied in various forms in the Jordan frame and the Einstein frame
while dealing with inflationary cosmology or structure
formation\cite{DeFelice:2010aj, Matsumoto:2013sba, Bertacca:2011wu,
  Mukhanov:1990me, Tsujikawa:2009ku}. In this section we will present
the Einstein frame description of the perturbations produced during a
cosmological bounce in the Jordan frame where the dynamics of the
perturbations is guided by a HD theory of gravity. As like the
background bouncing solution these perturbation can also be analyzed
in the Einstein frame. Once the solution of the metric perturbations,
corresponding to the bouncing solutions in the Jordan frame, are
solved in the Einstein frame one may transform the results back to the
Jordan frame.

Focussing only on the scalar perturbations one can write the perturbed
line element as, 
\begin{eqnarray}
ds^{2} = a^{2}\left(\eta \right)\left[-(1+2\phi)d\eta^{2} + 
2B_{,i}\,d\eta \,dx^{i}+\left\lbrace(1-2\psi)\delta_{ij}+2 E_{,ij}
\right\rbrace dx^{i}dx^{j}
\right]\,,\nonumber\\
\label{dsj}
\end{eqnarray}
where $ \phi,\psi,B,E$ characterize the scalar perturbations and
commas followed by $i$ and $j$ appearing after the functions $B$ and
$E$ refer to partial differentiation of the functions with respect to
$x^i$ or $x^j$. One can construct the gauge invariant Bardeen
potentials by suitable combinations of the scalar functions as
\begin{eqnarray}
\Phi=\phi+\frac{1}{a}[a(B-E^{'})]^{'}\,, \,\,\,\,\,\,\,\, 
\Psi=\psi-\frac{a^{'}}{a}(B-E^{'})\,.
\label{bpj}
\end{eqnarray}
It has to be emphasized at this point that the primes over the
quantities in this section will mean a derivative with respect to
the conformal time. The conformal time remains the same both in the
Jordan and the Einstein frames. In terms of gauge invariant Bardeen
potentials, the perturbed metric in the Jordan frame is given as
\begin{eqnarray}
ds^{2}=a^{2}(\eta)\left[ -(1+2\Phi)d\eta^{2}+(1-2\Psi)\delta_{ij} 
dx^{i}dx^{j}\right]\,,
\label{gidsj}
\end{eqnarray}
where $\Phi \ne \Psi$ in HD theories of gravity.  The
perturbed metric in the Einstein frame can also be written as
\begin{eqnarray}
d\tilde{s}^{2}=\tilde{a}^{2}(\eta)\left[-(1+2\tilde{\Phi})d\eta^{2}
+(1-2\tilde{\Psi})\delta_{ij} dx^{i}dx^{j}\right]\,,
\end{eqnarray}
As the energy momentum tensor, $\tilde{S}^\mu_\nu$, as given in
Eq.~(\ref{mttilde}) in the Einstein frame is diagonal one has
$\tilde{\Phi} = \tilde{\Psi}$.  The Bardeen potentials in the Jordan
frame can be calculated from $\tilde{\Phi}$ in the following way 
\cite{Mukhanov:1990me} :
\begin{eqnarray}
\Phi= -\frac{2}{3}\left(\frac{F^{2}}{F'a}\right)\left[\left(\frac{a}{F}
\right)\tilde{\Phi}\right]'\,,\,\,\,\,
\Psi = \frac{2}{3}\left(\frac{1}{FF'a}\right)(aF^{2}\tilde{\Phi})'\,.
\label{phipsi}
\end{eqnarray}
Assuming the perturbation in the matter sector to be adiabatic in the
Jordan frame, one can write
\begin{eqnarray}
\delta p^{(gi)}-c_{s}^{2}\delta \rho^{(gi)} =0\,.
\label{del}
\end{eqnarray}
For any quantity $q$, $\delta q^{(gi)} \equiv \delta q +
q^{\prime}_{0}(B-E^{\prime})$, where $q_{0}$ denotes the unperturbed
quantity.  Replacing $\rho, p$ by $\tilde{\rho},\tilde{p}$ in the
Eq.~(\ref{del}), and keeping in mind that ${\delta F}/{F} =
\sqrt{{2\kappa}/{3}}\,\delta\varphi$, where $\varphi(\eta,{\bf
  x})=\varphi_0(\eta) + \delta \varphi(\eta,{\bf x})$ one obtains
\begin{eqnarray}
\delta\tilde{p}^{(gi)}-c_{s}^{2}\delta\tilde{\rho}^{(gi)}=2\sqrt{\frac{2\kappa}{3}}
\tilde{\rho}_0(c_{s}^{2}-\omega)\delta\varphi^{(gi)}\,,
\end{eqnarray}
where $\tilde{\rho}_0$ is the background energy density of the
effective fluid.  Till the last section this parameter was not
represented with the zero subscript but in this section we use this
convention to differentiate the background values from their
perturbations. The above equation shows that if one has a single
component barotropic fluid for which $c_s^2=\omega$ in the Jordan
frame, the hydrodynamic perturbations produced in the Einstein frame
are also adiabatic. In terms of the Hubble parameter
$\tilde{\mathcal{H}}$, now represented as a function of the conformal
time, the gauge invariant perturbed field equations in the Einstein
frame are:
\begin{eqnarray}
 -3\tilde{\mathcal{H}}(\tilde{\mathcal{H}}\tilde{\Phi}+\tilde{\Phi}^{'})+
\nabla^{2}\tilde{\Phi} &=& \frac{\kappa}{2} 
\left( \tilde{a}^{2}\delta \tilde{\rho}^{(gi)}
-{\varphi^{\prime}_{0}}^{2}
\tilde{\Phi} +\varphi^{\prime}_{0}\delta\varphi^{(gi)^{\prime}}
 \right.  \nonumber \\
&& \left.  + \tilde{a}^{2}V_{,\varphi_0}
 \delta\varphi^{(gi)}\right)\,,
\label{pert1}\\ 
(2\tilde{\mathcal{H}}^{\prime}+\tilde{\mathcal{H}}^{2}) \tilde{\Phi} +
\tilde{\Phi}^{\prime \prime}
+3\tilde{\mathcal{H}}\tilde{\Phi}^{\prime} &=& \frac{\kappa}{2}
\left( \tilde{a}^{2}\delta
\tilde{p}^{(gi)} -
\tilde{\Phi} {\varphi^{\prime}_{0}}^{2} + \varphi^{\prime}_{0} 
\delta\varphi^{(gi)^{'}}\right. \nonumber\\
&& \left. -\tilde{a}^{2} V_{,\varphi_0}\delta\varphi^{(gi)}\right)\,,
\label{pert2}
\end{eqnarray}
where in the above equations $V_{,\varphi_0}$ specifies the derivative
of the scalar field potential with respect to $\varphi_0$.
Multiplying Eq.~(\ref{pert1}) by $c_s^2$ and subtracting the
resultant from Eq.~(\ref{pert2}) yields:
\begin{eqnarray}
& &\tilde{\Phi}^{\prime \prime}- c_{s}^{2}\nabla^{2}\tilde{\Phi} +
\left( 2\tilde{\mathcal{H}}^{\prime} + 
\tilde{\mathcal{H}}^{2}
  \right)\tilde{\Phi} + 
3 \tilde{\mathcal{H}}\tilde{\Phi}^{\prime} + 
3 c_{s}^{2}\tilde{\mathcal{H}}\left(  \tilde{\mathcal{H}}\tilde{\Phi} 
+ \tilde{\Phi}^{\prime} \right) 
= -\frac{\kappa}{2} \tilde{\Phi}\varphi^{\prime 2}_{0}
\left( 1- c_{s}^{2} \right)
\nonumber \\
&& +\frac{\kappa}{2}\varphi^{\prime}_{0}(1-c_{s}^{2})\delta
{\varphi^{(gi)}}^\prime - \frac{\kappa \tilde{a}^2}{2} V_{,\varphi_0}
(1 + c_{s}^{2})\delta \varphi^{(gi)}\,.
\label{intpert}
\end{eqnarray}
In the rest frame of the hydrodynamic fluid one can write
\begin{eqnarray}
\frac{\kappa}{2}\delta\varphi^{(gi)} &=&
\frac{1}{\varphi^{\prime}_{0}}
\left[\tilde{\Phi}^{'}
+\tilde{\mathcal{H}}\tilde{\Phi} \right]\,,
\end{eqnarray}
and taking it's derivative with respect to $\eta$ one gets
\begin{eqnarray}
\frac{\kappa}{2}\varphi^{\prime}_{0}\delta {\varphi^{(gi)}}^{\prime} &=& 
\left( \tilde{\mathcal{H}}^{\prime} - \frac{\varphi^{\prime
    \prime}_{0}}{\varphi^{\prime}_{0}}\tilde{\mathcal{H}}
\right)\tilde{\Phi} + \left( \tilde{\mathcal{H}}-
\frac{ \varphi^{\prime \prime}_{0}}{\varphi^{\prime}_{0}} \right) 
\tilde{\Phi}^{\prime} + \tilde{\Phi}^{\prime \prime}\,.
\end{eqnarray}
Using the above equation, the form of Eq.~(\ref{phiev}) written in
conformal time and
$$\frac{\kappa}{2} \varphi_0'^2 = \tilde{\mathcal{H}}^2 -
\tilde{\mathcal{H}}'- \frac{\kappa \tilde{a}^2}{2}(\tilde{\rho}_0 +
\tilde{p_0})\,,$$ which is obtained from the Einstein equations for
the background spacetime, in Eq.~(\ref{intpert}) we obtain the
equation for scalar perturbation in the Einstein frame as:
\begin{eqnarray}
&&c_s^2(\tilde{\Phi}^{''} - \nabla^{2}\tilde{\Phi}) + 
\left[2c_s^2\left(\tilde{\mathcal{H}}
-\frac{\varphi_{0}^{''}}{\varphi_{0}^{'}}\right)+
\frac{\tilde{a}^{2}}{\varphi_{0}^{'}}\sqrt{\frac{\kappa}{6}}\,\,\tilde{\rho}_0
(1+c_s^2)(1-3c_{s}^{2})\right]\tilde{\Phi}^{'}\nonumber\\
&+&\left[2\left(\tilde{\mathcal{H}}^{'}-\tilde{
\mathcal{H}}\frac{\varphi_{0}^{''}}{\varphi_{0}^{'}}\right)c_s^2
+ \tilde{a}^{2}\tilde{\rho}_0(1+c_s^2)\left\lbrace 
\sqrt{\frac{\kappa}{6}}\frac{\tilde{\mathcal{H}}}{\varphi_{0}^{'}}
(1-3c_{s}^{2})-\frac{\kappa}{2}
(1-c_{s}^{2})\right\rbrace \right]
\tilde{\Phi} = 0\,,\nonumber\\
\label{mpeqn}
\end{eqnarray}
where $\tilde{\rho}_0$ and $\tilde{p}_0$ are the background energy
density and pressure of the effective fluid.  For a single barotropic
fluid in the Jordan frame $\tilde{p}_0=c_s^2 \tilde{\rho}_0$ where
$c_s^2$ is the sound velocity in the Jordan frame. In absence of any
hydrodynamic fluid in the Jordan frame the above equation becomes
\begin{eqnarray}
\tilde{\Phi}^{''} - \nabla^{2}\tilde{\Phi} + 
2\left(\tilde{\mathcal{H}}
-\frac{\varphi_{0}^{''}}{\varphi_{0}^{'}}\right)\tilde{\Phi}^{'}
+\left[2\left(\tilde{\mathcal{H}}^{'}-\tilde{
\mathcal{H}}\frac{\varphi_{0}^{''}}{\varphi_{0}^{'}}\right) \right]
\tilde{\Phi} = 0\,,
\label{mpeqn1}
\end{eqnarray}
which is equivalent to the corresponding equation for scalar metric
perturbation in Ref.~\cite{Martin:2003sf} where the authors used the
background FRW solution to have a positive curvature. If the curvature
term in Ref.~\cite{Martin:2003sf} is dropped then the equation for the
scalar metric perturbation equation matches exactly with the above
equation \footnote{Actually Eq.~(\ref{mpeqn1}) is written for the case
  where the background FRW solution is spatially flat, $k=0$. In the
  general case, where $k\ne 0$ the coefficient function multiplying
  $\tilde{\Phi}$ in Eq.~(\ref{mpeqn1}) will have an extra piece, $-4k$,
  added. From this modified form of Eq.~(\ref{mpeqn1}) it can be
  easily verified that it reduces to the corresponding equation for
  the metric perturbation in Ref.~\cite{Martin:2003sf}. In
  Ref.~\cite{Martin:2003sf} as the authors used positively curved
  spatial hypersurfaces in FRW spacetime there they used
  $\nabla^{2}\tilde{\Phi}=-n(n+2)\tilde{\Phi}$ where $n$ is an
  integer.}.
 
If the background hydrodynamic fluid is made up of radiation then
Eq.~(\ref{mpeqn}) takes a simpler form as
\begin{eqnarray}
&&\tilde{\Phi}^{''} - \nabla^{2}\tilde{\Phi} + 
2\left(\tilde{\mathcal{H}}
-\frac{\varphi_{0}^{''}}{\varphi_{0}^{'}}\right)\tilde{\Phi}^{'}
+\left[2\left(\tilde{\mathcal{H}}^{'}-\tilde{
\mathcal{H}}\frac{\varphi_{0}^{''}}{\varphi_{0}^{'}}\right)
- \frac{4\kappa}{3}\tilde{a}^{2}\tilde{\rho}_0\right]
\tilde{\Phi} = 0\,,
\label{pertrad}
\end{eqnarray}
where we have used $c_s^2=1/3$. One can also use the background fluid
to be dust like and see the evolution of the perturbations across the
bounce.  Cosmological perturbations in presence of hydrodynamic matter
in general predicts that for a dust filled universe the perturbation
equations become a second order differential equation in time which is
independent of the wave number of the perturbations. This happens,
independently of the underlying theory of gravity, because the
wave number dependent terms in the perturbation equation come with a
factor of $c_s^2$ multiplied to it. In the present case, one may see
that the above argument is slightly modified. The form of
Eq.~(\ref{mpeqn}) predicts that in a dust filled universe the
differential equation predicting the evolution of the scalar
perturbation becomes a first order differential equation, in time,
which is independent of the wave number of the perturbation. From a
quantum mechanical point of view this poses a problem because in those
cases it is difficult to impose initial conditions on the
perturbations and so the general ansatz is to hold $c_s^2 \to 0$, not
exactly zero, such that some wave number dependence persists in the
theory. In the next subsection we will present the numerical evolution
of the, classical perturbations, in the Jordan frame for which the
sound velocity will be taken arbitrarily small but not exactly
zero. The issue of quantization of the perturbations will be dealt in
a future publication and we will not discuss about it in this article.

Before leaving the topic of cosmological perturbations in the Einstein
frame it is important to note one interesting property about the main
evolution equation Eq.~(\ref{mpeqn}). The general technique to solve
differential equations, like Eq.~(\ref{mpeqn}), is to expand
$\tilde{\Phi}(\eta, {\bf x})$ in the eigenfunction basis of operator
$\nabla^{2}$. In such cases the Laplacian operator is replaced by the
eigenvalue of the operator. Ultimately one can transform
Eq.~(\ref{mpeqn}) into a second order differential equation for
$\tilde{\Phi}(\eta)$ in which the coefficients of $\tilde{\Phi}^{'}$
and $\tilde{\Phi}$ will be functions of the form
$\varphi_{0}^{''}/\varphi_{0}^{'}$, $1/\varphi_{0}^{'}$ or ${\mathcal
  H}\varphi_{0}^{''}/\varphi_{0}^{'}$. In our analysis of the behavior
of the scalar field in the Einstein frame, in section \ref{einf}, it
was pointed out that during the bouncing process $\varphi_0'$ always
reaches zero at some time and that is the turning point for
$\varphi_0$. From the form of Eq.~(\ref{mpeqn}) it seems that
coefficients in the differential equation becomes singular at the
value of $\eta$ when $\varphi_0'=0$. But under closer inspection it
becomes clear that these singular points are actually regular singular
points of the equation. In general if $\varphi_0'(\eta_0)=0$ it can be
easily seen that
$\lim_{\eta\to\eta_0}((\eta-\eta_0)g(\eta)/\varphi_{0}^{'})$, or
$\lim_{\eta\to\eta_0}((\eta-\eta_0)^2 g(\eta)/\varphi_{0}^{'})$ where
$g(\eta)$ is some function of $\eta$, is always nonsingular if
$\varphi_0''(\eta_0) \ne 0$. One must always have
$\varphi_0''(\eta_0) \ne 0$ as $\eta_0$ is a turning point of
$\varphi_0$ where the rate of change of of $\varphi_0'$ is non
zero. These fact implies that there must exist well behaved solutions
of the cosmological perturbation evolution equation throughout the
bouncing regime.
%%%%%%%%%%%%%%%%%%%%%%%%%%%%%%%%%%%%%%%%%%%%%%%%%%%%%%%%%%%%%%%%%%%%%%%%%%%
\subsection{Numerical evolution of the scalar metric perturbations
  across the bounce}
In this subsection we present the numerical evolution of the growth of
the scalar cosmological perturbations in radiation dominated and
matter dominated universes whose cosmology is governed by quadratic
$f(R)$ gravity during a brief bouncing period centered at $\eta=0$.
The results yield plots of $\Phi$ and $\Psi$ in the Jordan frame for
specific values of the wave number $k$ of the perturbations. To
calculate the perturbations in the Jordan frame one has to use the
relations, as given in Eq.~(\ref{phipsi}), which connect
$\tilde{\Phi}$ in the Einstein frame to the corresponding perturbation
potentials in the Jordan frame. The relations in Eq.~(\ref{phipsi})
reveals that the Jordan frame perturbation potentials can diverge when
$F'=0$. It is known that $F'$ must vanish at one instant during the
bouncing period.  Mathematically one cannot use the Einstein frame
description of the perturbations to track the development of the
metric perturbations in the Jordan frame completely. For an asymmetric bounce,
when $F'=0$ at an instant, the Jordan frame potentials will become
singular when they are calculated using Eq.~(\ref{phipsi}). The
singularity of the Jordan frame metric perturbations at a particular
instant of time is purely an artefact of the transformation relations
of the perturbations in different conformal frames as given in
Eq.~(\ref{phipsi}) and does not represent a real singularity of the
perturbations as no such singularity in the evolution of the metric
perturbation $\tilde{\Phi}$ appears in the Einstein frame.  The
singularities in the relations in Eq.~(\ref{phipsi}) do not arise for
the symmetric bounce cases when the perturbation modes are also
symmetric in time in the Einstein frame. In this case $F'= a'=
\tilde{\Phi}'=0$ at $\eta=0$ and one can easily verify that the
connecting relations in the perturbations remain well behaved
throughout the bouncing phase. For an asymmetric perturbation mode in
the case of a symmetric background evolution one can have
\begin{figure}[t!]
\centering
\includegraphics[scale=.8]{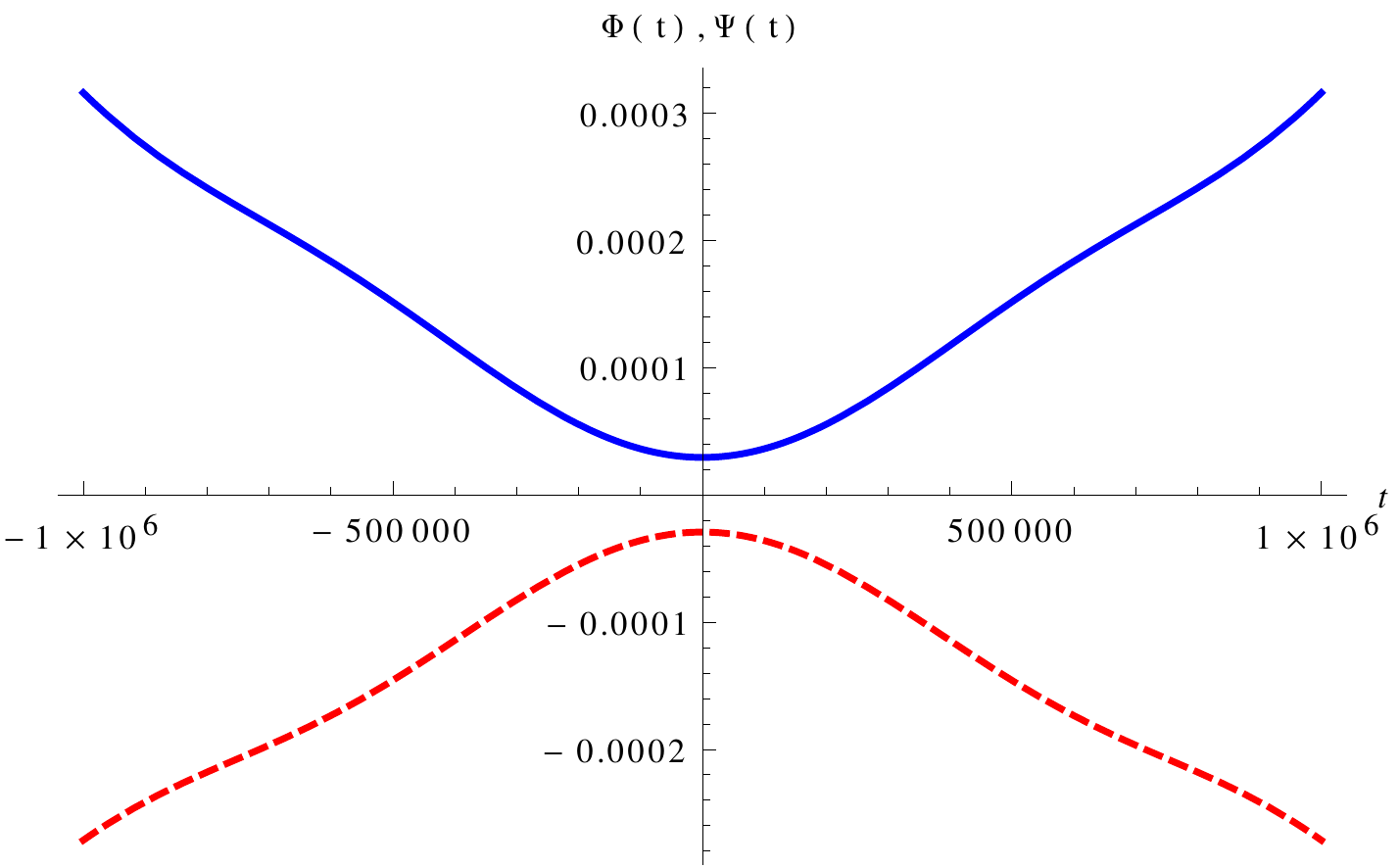}
\caption{The plot showing the evolution of $\Phi$ and $\Psi$ in the
  Jordan frame for radiation induced cosmological bounce. The blue
  coloured solid curve represents $\Phi$ and the red dashed curve
  represents $\Psi$. The perturbations are plotted with respect to
  normal Jordan frame time variable $t$. }
\label{rad_pert}
\end{figure}
$\tilde{\Phi}'\ne 0$ at $\eta=0$ and in such a case the discontinuity
of the Jordan frame perturbation potentials can reappear.  The above
discussion shows that the Einstein frame description of the evolution
of the perturbations can be used to study the perturbations in the HD
theory as the divergence of the scalar functions in the Jordan frame
(if present) are purely local (about an instant).  In this subsection
we will be presenting the perturbation evolutions for the symmetric
bounces as well as the asymmetric bounces for various kinds of
hydrodynamic fluid in the Jordan frame.

At first we study the symmetric bounce case where the perturbation
modes are also symmetric in time. In this case it is better to specify
the initial conditions for the numerical evolution of the
perturbations at $\eta=0$.  In Fig.~\ref{rad_pert} we show the
evolution of the scalar metric perturbations in the Jordan frame where
the bounce is triggered by a radiation fluid. The wave number for the
perturbation mode, $k=10^{-12}$ , correspond to a superhorizon mode at
$t=-10^{6}$.  Although the perturbation calculations were done using
conformal time we represent the final evolutions of the perturbations
in the Jordan frame using the Jordan frame time variable $t$. To
generate the numerical solutions we have assumed that the bounce is
symmetric, where for the background evolution $\phi(0)=-0.15$,
$\phi'(0)=0$ and for the perturbations $\tilde{\Phi}(0)=10^{-8}$ and
$\tilde{\Phi}'(0)=0$ in the Einstein frame. The magnitude of
$\tilde{\Phi}(0)$ in the Einstein frame is chosen in such a fashion
that the magnitude of the metric perturbations in the Jordan frame are
of the order of $10^{-4}$.  From Fig.~\ref{rad_pert} it is seen that
the quantities $\Phi$ and $\Psi$ practically remains constant during
the bouncing period, although a slight variation is noted in their
magnitudes. The evolution of the perturbations also remain symmetric
in time.
\begin{figure}[t!]
\centering
\includegraphics[scale=1.1]{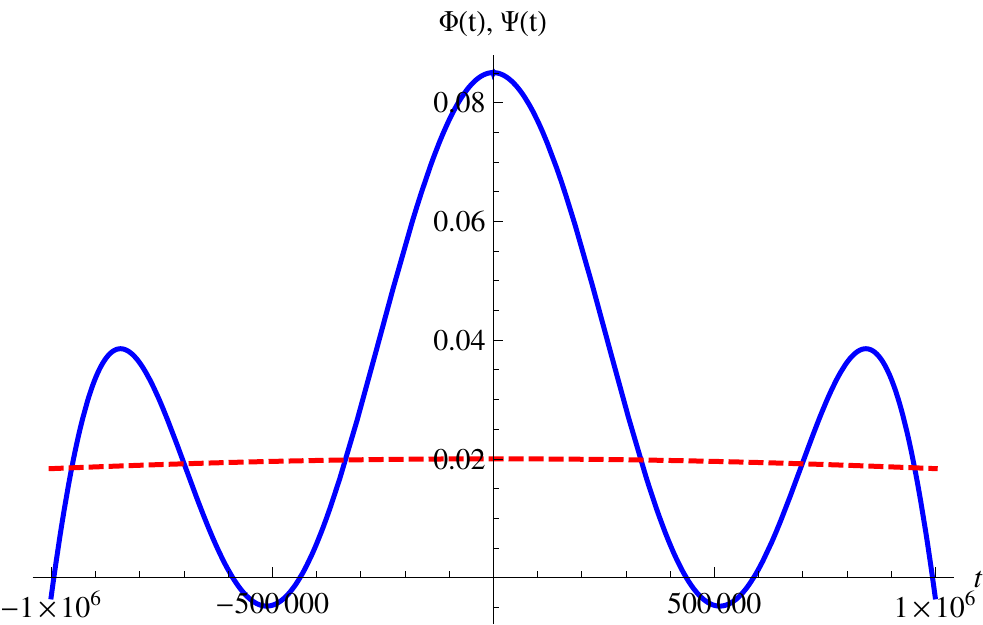}
\caption{The plot showing the evolution of $\Phi$ and $\Psi$ in the
  Jordan frame for dust matter induced bounce. The blue
  coloured solid curve represents $10^{4} \times \Phi$, and the red
  dashed curve represents $\Psi$.}
\label{dust_pert}
\end{figure}

The corresponding perturbation evolution in the dust dominated
bouncing universe is shown in Fig.~\ref{dust_pert}. In this case also
$k=10^{-12}$, for the perturbation mode, which corresponds to a
superhorizon mode in the beginning phase of the bouncing period. To
keep the $k$ dependent terms in the perturbation evolution we have
used $c_s^2=10^{-8}$ in Eq.~(\ref{mpeqn}). In this case also it is noticed
that the symmetry of the bounce is reflected in the perturbation
modes. Although $\Phi$ oscillates in the bouncing period but its
overall variation in the that period is negligible. The perturbations
remain practically constant through the bouncing phase. In this case
initial values used to determine the perturbations in the Einstein frame are,
$\tilde{\Phi}(0)=10^{-2}$ and $\tilde{\Phi}'(0)=0$ and the background
values of $\phi(0)$ and $\phi'(0)$ remain the same as in the case of
radiation induced bounce.

\begin{figure}[t!]
\begin{minipage}[b]{0.5\linewidth}
\centering \includegraphics[scale=.7]{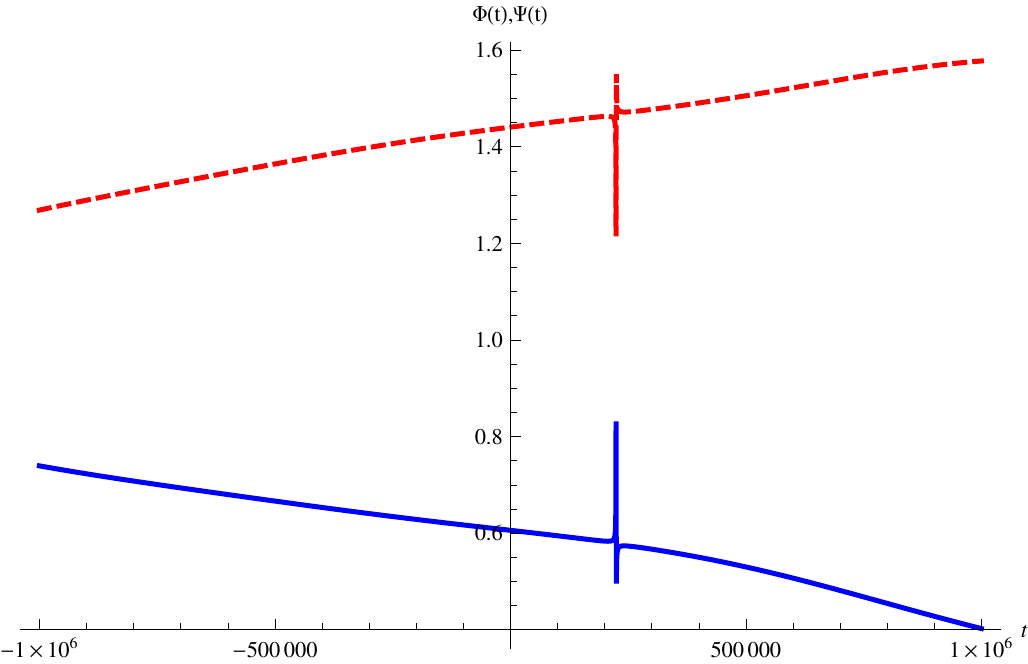}
\caption{Metric perturbation potentials, $\Phi$ (solid blue) and $\Psi$
  (dashed red), in the Jordan frame for an
  asymmetric radiation induced bounce where $\tilde{\Phi}=1$, and
  $\tilde{\Phi}'=0$ at $\tilde{t}=-10^6$.}
\label{rad7}
\end{minipage}
\hspace{0.2cm}
\begin{minipage}[b]{0.5\linewidth}
\centering
\includegraphics[scale=.7]{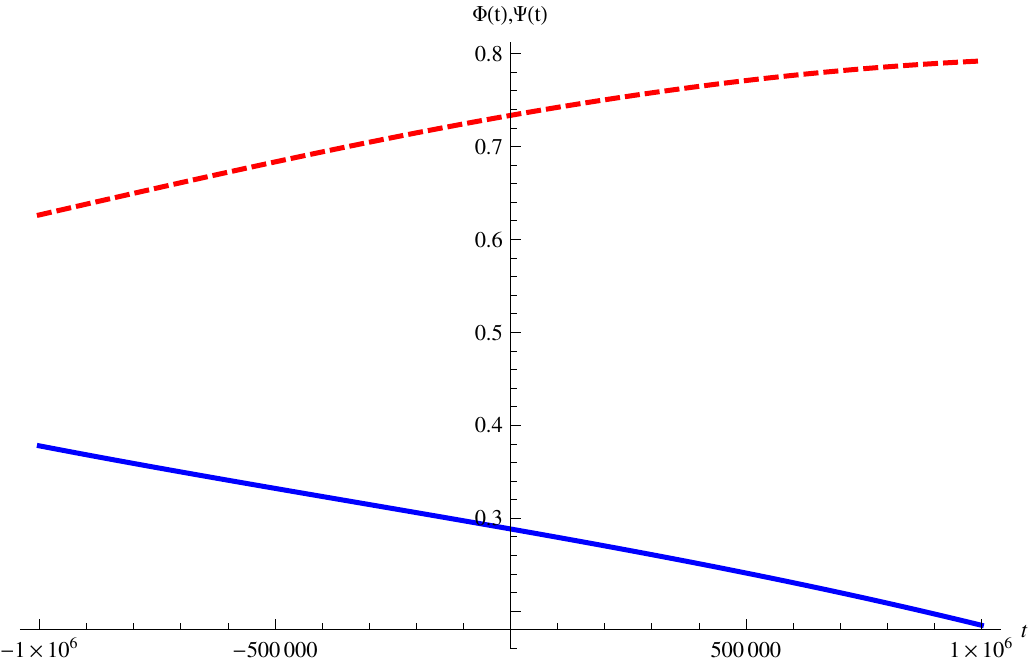}
\caption{Metric perturbation potentials, $\Phi$ (solid blue) and $\Psi$
  (dashed red), in the Jordan frame for 
  symmetric radiation induced bounce where $\tilde{\Phi}=0.5$, and
  $\tilde{\Phi}'=0$ at $\tilde{t}=-10^{6}$.}
\label{rad10}
\end{minipage}
\end{figure}
Next we present the more general results for both symmetric and
asymmetric bounces where the perturbations may not be symmetric. In
this cases we set the initial conditions for the perturbation at
$\tilde{t}=-10^6$ where time is expressed in Planck units.  In
Figs.~\ref{rad7} and \ref{rad10} we present the perturbation
evolutions for an asymmetric and symmetric radiation induced bounce in
quadratic gravity. Here we immediately see that in Fig.~\ref{rad7}
there is a sharp jump of the metric perturbations near the bouncing
point and one can verify that at that time $F'(R)=0$. The jumps are
shown to be finite in the plots because of limitation of the numerical
approach due to which the exact value of $F'(R)=0$ is never reached.
But Fig.~\ref{rad7} shows that except one single point the evolution
of the perturbations in the Jordan frame is perfectly well defined for
an asymmetric bounce. The background evolution for the asymmetric
bounce cases discussed here and henceforth will use the initial
conditions as specified in Eq.~\ref{initphip} and the symmetric bounce
initial conditions remain the same as discussed before in this
subsection. The wave numbers for the modes will we chosen to be
$k=10^{-12}$ throughout. The perturbations are seen to be nearly
constant in the Jordan frame. Interestingly we do not observe the
discontinuities of $\Phi$ and $\Psi$ at $t=0$ in the symmetric
bounce. They should in principle be present as now the initial
conditions for the perturbation evolutions do not ensure that
$\tilde{\Phi}'=0$ at $\eta=0$. They remain unobservable because
the amplitude of the discontinuous jump at $t=0$ for both the
perturbation potentials turns out to be much smaller than the original
functional values of these variables at $t=0$ and consequently they
remain hidden and the perturbation potentials look smooth at $t=0$ in
the Jordan frame. We will show that in the matter induced bounce
background one will be able to observe these discontinuous jumps in
$\Phi$ even for a symmetric bounce background as there the amplitude
of the perturbing potential $\Phi$ is itself very small and the
amplitude of the ``apparent'' finite discontinuity does not remain
hidden.
\begin{figure}[t!]
\begin{minipage}[b]{0.5\linewidth}
\centering \includegraphics[scale=.7]{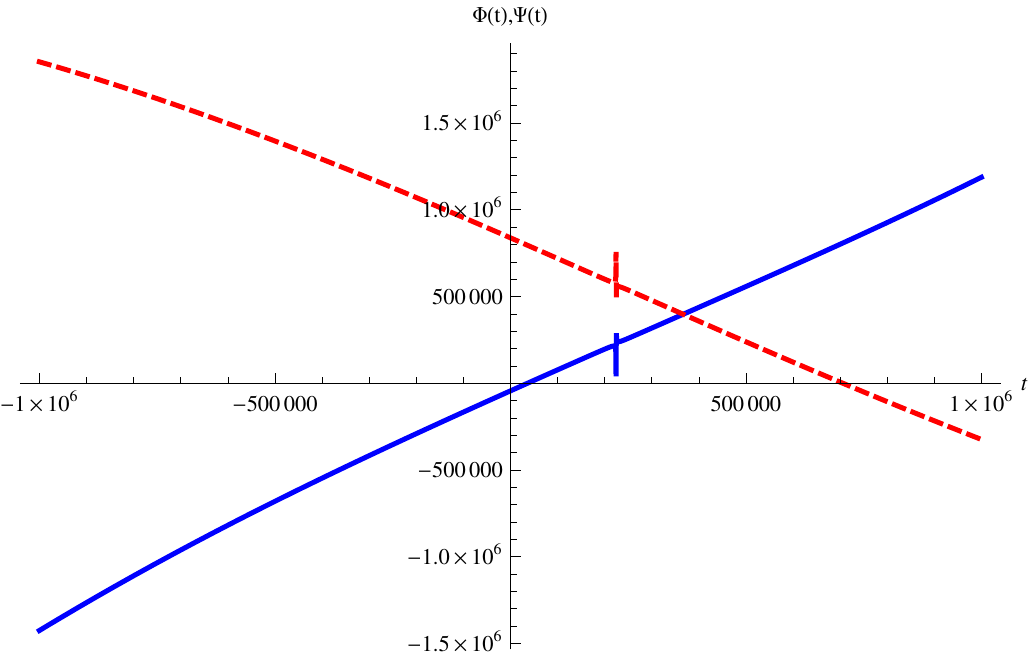}
\caption{Metric perturbation potentials, $\Phi$ (solid blue) and $\Psi$
  (dashed red), in the Jordan frame for an
  asymmetric radiation induced bounce where $\tilde{\Phi}=0$, and
  $\tilde{\Phi}'=1$ at $\tilde{t}=-10^6$.}
\label{rad5}
\end{minipage}
\hspace{0.2cm}
\begin{minipage}[b]{0.5\linewidth}
\centering
\includegraphics[scale=.7]{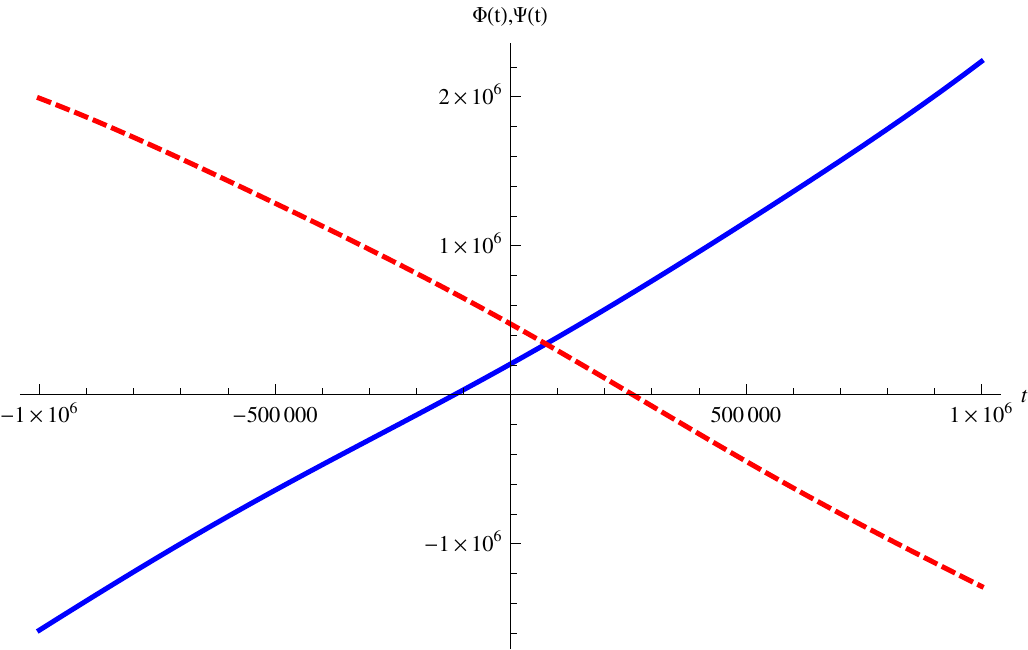}
\caption{Metric perturbation potentials, $\Phi$ (solid blue) and $\Psi$
  (dashed red), in the Jordan frame for 
  symmetric radiation induced bounce where $\tilde{\Phi}=0$, and
  $\tilde{\Phi}'=1$ at $\tilde{t}=-10^6$.}
\label{rad9}
\end{minipage}
\end{figure}

All the initial conditions set in the Einstein frame do not translate
into well behaved constant perturbations in the Jordan frame. Our
numerical results show that the results are particularly sensitive to
the initial value of $\tilde{\Phi}'$. For large values of
$\tilde{\Phi}'$ as initial condition one gets the perturbing
potentials $\Phi$ and $\Psi$ to be very large and the theory does not
remain perturbative.  This fact is shown in the specimen plots in
Figs.~\ref{rad5} and \ref{rad9} where the values of the perturbing
potentials are practically diverging throughout the bouncing
regime. The result is not surprising as if one sets $\tilde{\Phi}'=1$
at the initial stage then the perturbation potential $\tilde{\Phi}$ in
the Einstein frame is steeply increasing with time which may attain a
very high value very quickly. The transformed Jordan frame
perturbation potentials reflect the unboundedness of the perturbation
in the Einstein frame. The plots in Fig.~\ref{rad5} and
Figs.~\ref{rad9} and the discussion above shows that the interesting
initial conditions, in the Einstein frame, which give rise to
physically reliable scalar metric perturbation evolution in the Jordan
frame must reject high values of $\tilde{\Phi}'$. On the other hand
our numerical evolutions show that the results are also dependent on
the magnitude of $\tilde{\Phi}$ initially. If $\tilde{\Phi} \sim 1$
then the perturbation potentials are approximately near to one, as in
Fig.~\ref{rad7}, but may grow as time increases in some other
cases. In the perturbative regime of the potential, $\tilde{\Phi} \lesssim
1$, and its derivative $\tilde{\Phi}' \lesssim 1$ the growth of $\Phi$ and
$\Psi$ in the Jordan frame remain well behaved.

\begin{figure}[t!]
\begin{minipage}[b]{0.5\linewidth}
\centering \includegraphics[scale=.75]{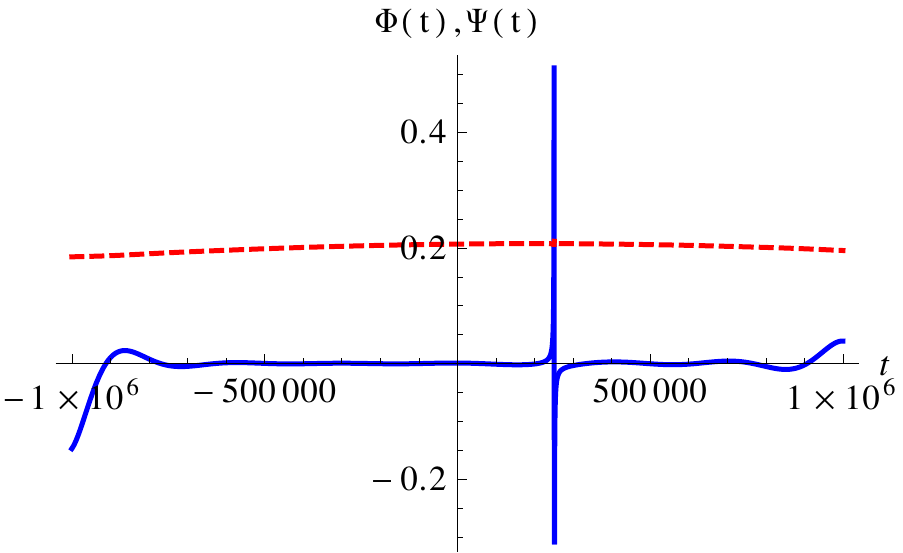}
\caption{Metric perturbation potentials, $\Phi \times 10^2$ (solid blue) and $\Psi$
  (dashed red), in the Jordan frame for an
  asymmetric dust matter induced bounce where $\tilde{\Phi}=0$, and
  $\tilde{\Phi}'=1$ at $\tilde{t}=-10^6$.}
\label{mat1}
\end{minipage}
\hspace{0.2cm}
\begin{minipage}[b]{0.5\linewidth}
\centering
\includegraphics[scale=.5]{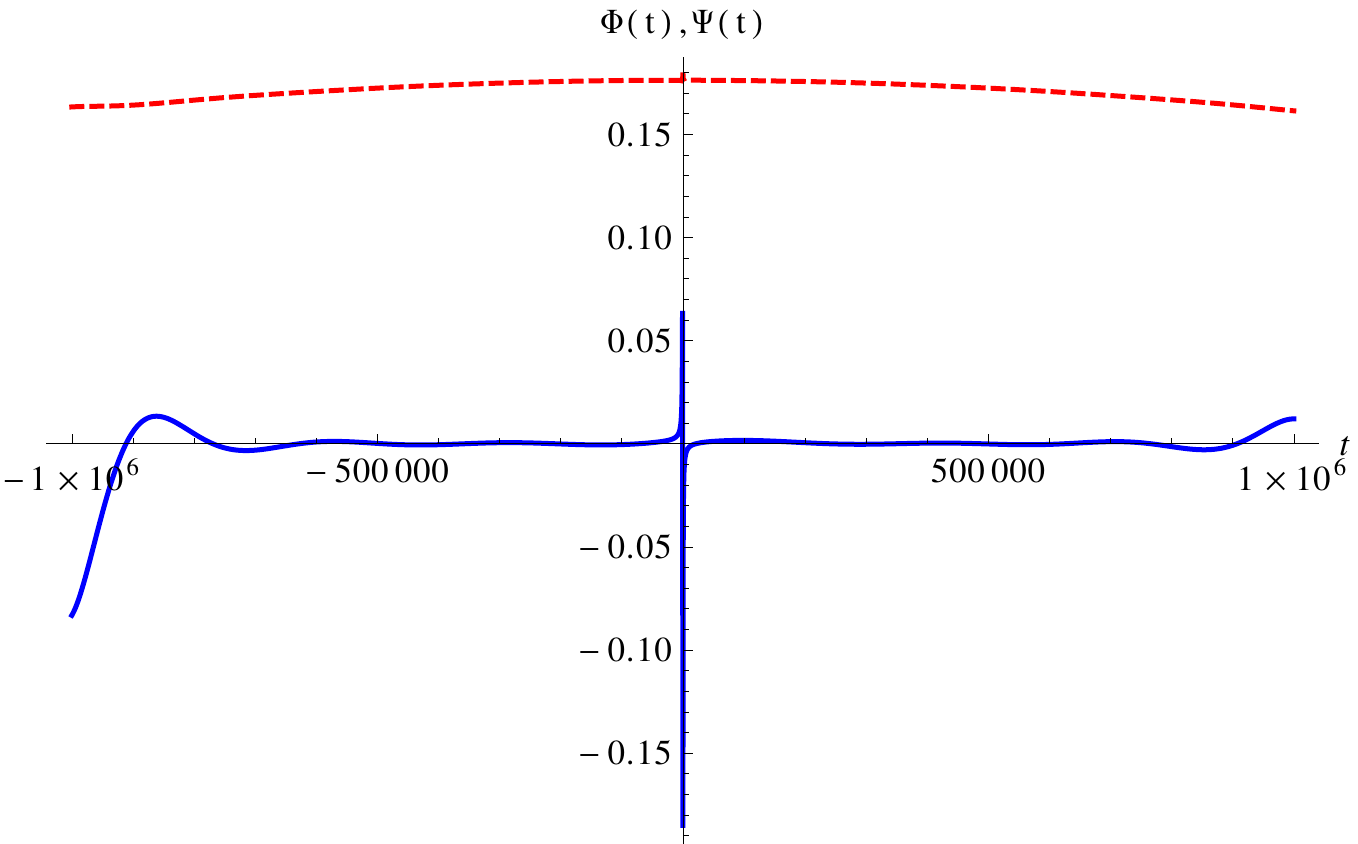}
\caption{Metric perturbation potentials, $\Phi \times 50$ (solid blue)
  and $\Psi$ (dashed red), in the Jordan frame for 
  symmetric dust matter induced bounce where $\tilde{\Phi}=0$, and
  $\tilde{\Phi}'=1$ at $\tilde{t}=-10^6$.}
\label{mat4}
\end{minipage}
\end{figure}

The pattern of time evolution of the perturbation potentials for the
dust induced bounce is similar to the radiation induced bounce except
a couple of points. We had commented earlier that for a matter induced
bounce the perturbation evolution equation in the Einstein frame
becomes more like a first order differential equation in
$\tilde{\Phi}$.  For the sake of the quantum mechanical initial
conditions it was assumed that for dust matter perturbations it is
safe to take $c_s^2 \to 0$ (where $c_s^2 \ne 0$) and consequently in
our plots we have always assumed $c_s^2 = 10^{-8}$. In such cases it
can be show that the effect of the second derivative of $\tilde{\Phi}$
with respect to the conformal time in Eq.~(\ref{mpeqn}) becomes
negligible and consequently the perturbation evolution equation
behaves more like a first order differential equation in
$\tilde{\Phi}$ which is heavily sensitive to the initial condition on
$\tilde{\Phi}$ ($\tilde{\Phi}(\eta=-10^6)$) and less sensitive on the
other initial condition on $\tilde{\Phi}'$. Consequently in this case
the perturbation evolution equations do not have large amplitudes as
one takes $\tilde{\Phi}' \to 1$. This fact becomes apparent
from the plots presented in Fig.~\ref{mat1} and Fig.~\ref{mat4} where
in the two plots $\tilde{\Phi}$ remains zero but $\tilde{\Phi}'=1$. In
this plots the initial conditions for background evolutions, for the
symmetric and asymmetric bounces respectively, remain the same as
those used for the radiation induced bounces. The wave number of the
perturbation mode is $k=10^{-12}$.

The other important point regarding the evolution of the perturbations
in the matter induced bounce case is related to presence of the
discontinuous jump in $\Phi$ in the symmetric bounce background. In
the radiation induced symmetric bounce backgrounds this jump was not
observed because the amplitude of the discontinuity was much less than
the numerical value of $\Phi$ or $\Psi$. In the matter induced bounce
the amplitude of $\Phi$ is much less (compared to the amplitude of
$\Phi$ in radiation induced bounce) and now the amplitude of the
discontinuity shows up. Interestingly in the present case it appears
that $\Psi$ is continuous in all the cases. Actually $\Psi$ also has
discontinuities at points where $F'=0$ but the amplitude of the
discontinuity is much less than the value of $\Psi$ and consequently
the discontinuities remain hidden.
%%%%%%%%%%%%%%%%%%%%%%%%%%%%%%%%%%%%%%%%%%%%%%%%%%%%%
\section{Discussion and Conclusion}
\label{conc}
As mentioned in the introduction of this article there are several
papers on cosmological bounce in the framework of GR and some on HD
gravity as well. In the works with HD gravity, mainly the bouncing
conditions with various forms of $f(R)$ were discussed. A thorough
analysis of the whole bouncing mechanism in $f(R)$ theories did not
come into sight of the present authors. More over the previous works
on cosmological bounce in $f(R)$ theories were done in the Jordan
frame itself where interpretation of the dynamics becomes too
complicated because of the HD terms involving the Ricci scalar $R$. In
the present article our attempt has been to understand the bouncing
cosmologies in $f(R)$ theories by studying the corresponding
cosmological behavior in the Einstein frame where the theory of
gravity is guided by general relativity. Although this prescription is
not foolproof as the corresponding theory in the Einstein frame can
come with a scalar field whose potential is not bounded from below,
but still one can safely use the Einstein frame description of the
bouncing phenomenon because before the scalar field starts to role
over to the infinite negative depth of the potential the HD theory of
gravity itself becomes ineffective.

In this article we started with the general bouncing conditions in the
Jordan frame and then focussed on FRW spacetimes which have flat
spatial hypersurfaces. A very general proof of the sufficient
conditions for the possibility of cosmological bounces, in the absence
of any hydrodynamic matter, in such kind of spacetimes is presented in
subsection \ref{mlb}. In this article we have worked out the full
Einstein frame formulation of the cosmological bounce.  The Jordan
frame and Einstein frame relationships for $f(R)$ theories for
inflationary cosmologies were known for a long time. But the present
case is different because unlike inflationary cosmology, where there
is an accelerated expansion of the universe in both the conformally
connected frames, here the two frames may show completely
different behaviors. We have always assumed that the physical frame is
the Jordan frame and so this divergent behaviors do not pose a
paradoxical result.

Specifically, the present article deals with the issue of studying
bouncing cosmologies in quadratic $f(R)$ theory of gravity where the
background FRW line element has zero spatial curvature. The bounces in
these cosmologies are interesting because the description of the
bouncing phenomena is very different in the Jordan frame and the
Einstein frame. If there are cosmological bounces in the Jordan frame
there cannot be any analogous bounce in the Einstein frame as shown in
subsection \ref{abef}. In the Einstein frame the bouncing regime
corresponds to an expanding phase followed by a contracting phase. The
change over time from an expanding to a contracting phase depends on
the values of the scalar field and its first derivative with respect
to the time variable in the Einstein frame.  Whereas for a FRW
spacetime with non flat spatial hypersurface one may get simultaneous
bounces in both the conformally related frames only when the curvature
of the spatial hypersurface of the FRW spacetime is positive. In this
article we have worked out the explicit nature of the bounces for an
universe filled with radiation. From the dynamics of the system in the
Einstein frame we could predict about the symmetries of the bouncing
phenomenon.

The other important issue discussed in this article is related to the
theory of cosmological perturbations in the bouncing universe whose
dynamics is guided by a $f(R)$ theory of gravity in the Jordan
frame. In this article we have only worked with the scalar sector of
the perturbations.  Like the background analysis in this case also we
first write down the perturbation equations in the Einstein frame,
where the analysis becomes relatively simple because the two Bardeen
potentials are identical. In this article we derive the main evolution
equation for the Bardeen potential in the Einstein frame where the
background cosmological evolution takes place in presence of a
hydrodynamic fluid and a scalar field. We have assumed that there is
only one barotropic fluid in the Jordan frame such that the
hydrodynamic perturbations in both the conformally related frames
remain adiabatic. The main equation specifying the evolution of the
perturbations in the Einstein frame is interesting because from its
structure one can see that both the second order time derivative and
the Laplacian of the Bardeen potential is multiplied by the square of
the speed of sound. Consequently, if the cosmological bounce occurs in
presence of dust then the evolution of the perturbation equations turn
out to be a first order differential equation in conformal time. The
main evolution equation of the perturbations reduces to a previously
well known evolution equation of the Bardeen potential in the absence
of any hydrodynamic fluid. The only difficulty with the present method
is related to the fact that the Einstein frame evolution of the
perturbation may not give the complete evolution of the Jordan frame
perturbation evolution as the Jordan frame potentials may diverge at
one instant of time during the bounce. It was shown that this is an
artefact of the transformation relations which connect the scalar
perturbation potentials in the two conformal frames. This difficulty
does not limit the scope of Einstein frame calculation, of the
perturbations, as the divergence of the Jordan frame perturbations is
purely a local phenomenon happening at an instant of time and the
Jordan frame potentials are well behaved throughout the bouncing
regime except at that particular time instant.

The way $c_s^2$ appears in Eq.~(\ref{mpeqn}) is in general different
from the way it appears in conventional evolution equation of the
Bardeen potential in the presence of hydrodynamic matter, where
$c_s^2$ only multiplies the Laplacian operator acting on the Bardeen
potential and there is no $c_s^2$ factor before the second order time
derivative of the potential. But it can be verified from the form of
Eq.~(\ref{mpeqn}) that precisely the positions of the square of the
sound velocity there makes the evolution equation correspond to the
well known evolution equation of the Bardeen potential in absence of
hydrodynamic matter and in the presence of a scalar field, as found in
Ref.~\cite{Martin:2003sf}, in appropriate limits.

In this article we did not get the scope to discuss the very important
issue of tensor perturbations in the bouncing universe model with
$f(R)$ gravity. We presented a numerical solution for the scalar
perturbations $\Phi$ and $\Psi$ in the last section which shows that
the perturbations remain practically constant during the bouncing
period. The topic of tensor perturbations and quantization of the
spacetime fluctuations will be addressed in a future publication.
%%%%%%%%%%%%%%%%%%%%%%%%%%%%%%%%%%%%%%%%%%%%%%%%%%%%%%%%%%%%%%%%%%%%%%%

%%%%%%%%%%%%%%%%%%%%%%%%%%%%%%%%%%%%%%%%
\end{document}